\documentclass[twocolumn,prx,superscriptaddress,showpacs]{revtex4-1}
\usepackage{amsmath,amssymb}
\usepackage{graphicx}
\usepackage{subfigure}
\usepackage{verbatim}
\usepackage{amsfonts}
\usepackage{dcolumn}
\usepackage{bm}
\usepackage{color}
\usepackage[colorlinks,citecolor=blue]{hyperref}

\usepackage{mathrsfs}

\newcommand{\beq}{\begin{eqnarray} }
\newcommand{\eeq}{\end{eqnarray} }
\newcommand{\Beq}{\begin{eqnarray*} }
\newcommand{\Eeq}{\end{eqnarray*} }
\newcommand{\Bmat}{\left(\begin{matrix}}
\newcommand{\Emat}{\end{matrix}\right)}
\newcommand{\up}{\uparrow}
\newcommand{\dn}{\downarrow}

\begin{document}

\title{Dirac and Chiral Quantum Spin Liquids on the Honeycomb Lattice in 
a Magnetic Field}

\author{Zheng-Xin Liu}
\affiliation{Department of Physics, Renmin University of China, Beijing 
100872, China}

\author{B. Normand}
\affiliation{Neutrons and Muons Research Division, Paul Scherrer Institute, 
CH-5232 Villigen-PSI, Switzerland}

\begin{abstract}
Motivated by recent experimental observations in $\alpha$-RuCl$_3$, we study 
the $K$-$\Gamma$ model on the honeycomb lattice in an external magnetic 
field. By a slave-particle representation and Variational Monte Carlo 
calculations, we reproduce the phase transition from zigzag magnetic order 
to a field-induced disordered phase. The nature of this state depends 
crucially on the field orientation. For particular field directions in the 
honeycomb plane, we find a gapless Dirac spin liquid, in agreement with recent 
experiments on $\alpha$-RuCl$_3$. For a range of out-of-plane fields, we 
predict the existence of a Kalmeyer-Laughlin-type chiral spin liquid, which 
would show an integer-quantized thermal Hall effect. 
\end{abstract}

\maketitle

The Kitaev model on the honeycomb lattice \cite{Kitaev2006} is exactly 
solvable and thus presents a fundamental paradigm for both gapped and gapless 
quantum spin liquids (QSLs). An applied magnetic field can turn the gapless 
phase into a gapped, non-Abelian QSL, which would have direct applications 
in topological quantum computation \cite{Nayak_RMP}. Although Kitaev-type 
interactions are realized in layered honeycomb-lattice materials, such as 
Na$_2$IrO$_3$ \cite{NaIrO_L09, NaIrO_L10, NaIrO_B11, NaIrO_B12, NaIrO_L12} 
and $\alpha$-RuCl$_3$ \cite{RuCl_B96, RuCl_B14, RuCl_B15}, their magnetically 
ordered ground states \cite{NaIrO_B12, NaIrO_L12, Fletcher_JCSA_1967,
Sears_PRB_91, Johnson_PRB_92, Cao_PRB_93}, preclude a QSL and indicate 
significant non-Kitaev interactions \cite{NaIrO_L10,KJGamma_L14}. 
Nevertheless, experimental observations of a continuum by inelastic neutron 
scattering (INS) \cite{proximate16} and a gapless mode at intermediate fields 
by thermal conductivity \cite{Ru_FieldThermal_L17} have been taken as evidence 
for the proximity of $\alpha$-RuCl$_3$ to Kitaev physics. 

Recent experiments have established that magnetic order 
in $\alpha$-RuCl$_3$ is weak and can be suppressed both by a magnetic field 
\cite{Ru_FieldThermal_L17, Ru_FieldNMR_L17, Ru_FieldNMR_arX17} and by pressure 
\cite{Ru_Presstrans_arX17, Ru_PressNMR_arX17}. The critical field, when applied 
in the honeycomb ($ab$) plane, is $B_c = 7.5$ T \cite{Ru_FieldThermal_L17, 
Ru_FieldNMR_arX17}, which is far below the saturation field \cite{Johnson_PRB}, 
and the resulting partially polarized but magnetically disordered state has 
been claimed to be a QSL. Numerous very recent studies of this phase are 
divided as to its nature, with nuclear magnetic resonance (NMR) in an 
out-of-plane field \cite{Ru_FieldNMR_L17}, specific heat \cite{Sears_PRB, 
Wolter_PRB}, and neutron spectroscopy \cite{Ru_Field_neutron_arX17} reporting 
a gapped QSL, whereas power-law temperature dependences observed by thermal 
conductivity \cite{Ru_FieldThermal_L17} and NMR in an in-plane field 
\cite{Ru_FieldNMR_arX17} suggest a gapless (nodal) QSL. NMR is a particularly 
sensitive probe of low-energy spin excitations and the spin-lattice relaxation 
rate, $1/T_1 \propto T^3$, observed \cite{Ru_FieldNMR_arX17} at $T < 10$ K 
over a finite field range 7.5 T $< B <$ 12 T matches precisely the result 
anticipated for the point-node dispersion of a generic Kitaev system 
\cite{GaplessRespons_Balents_L16}, albeit only at zero field. Because 
such nodal excitations are neither magnons nor the Majorana fermions of 
the Kitaev QSL, both of which show a gapped spin response in a field 
\cite{SpinGapinKSL_L14}, their existence would pose a theoretical challenge. 
Because their density of states vanishes as energy approaches zero, their 
detection and distinction from a fully gapped phase poses a subtle 
experimental challenge.

\begin{figure}[t]
\includegraphics[width=4.4cm]{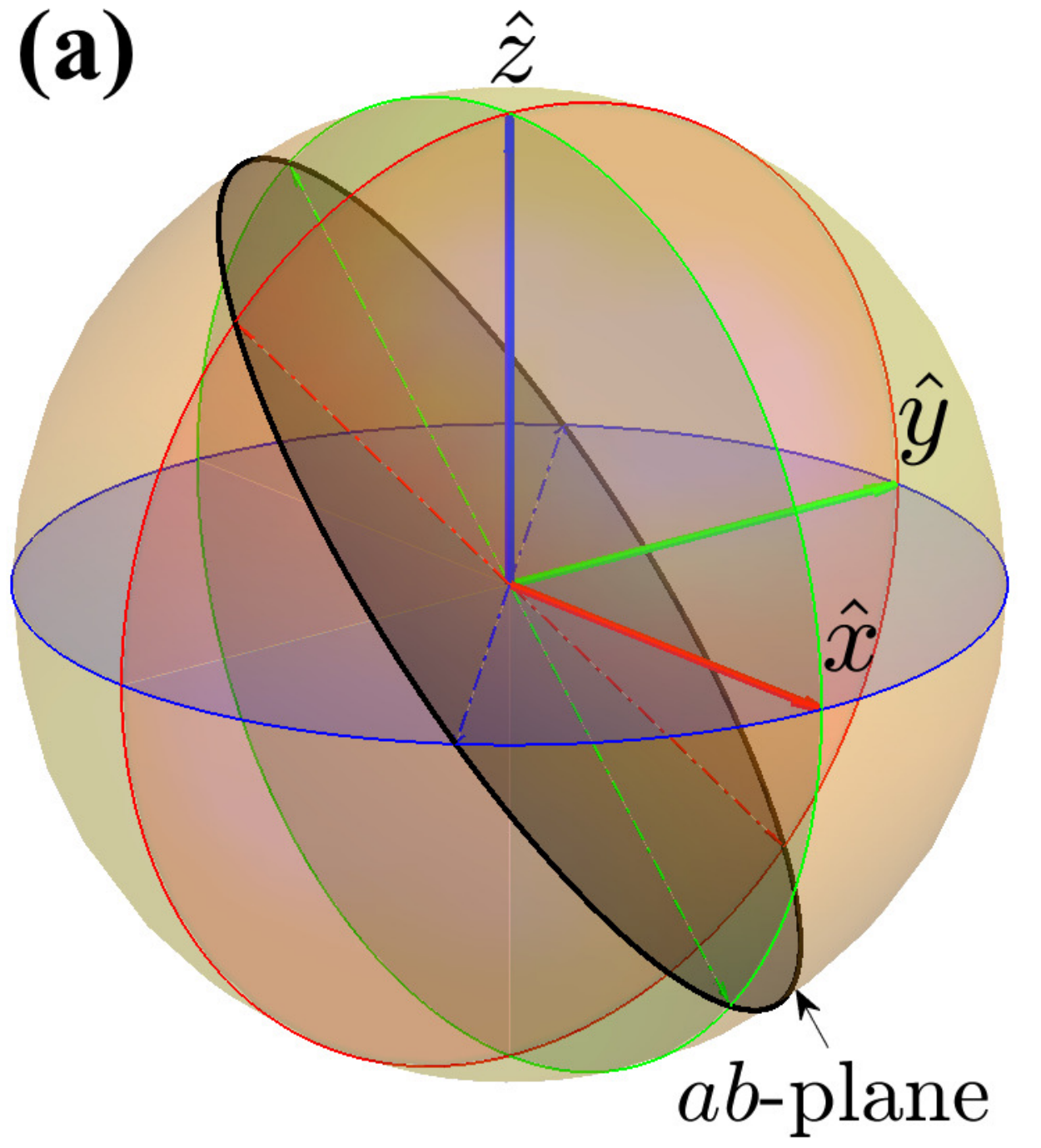}
\includegraphics[width=3.6cm]{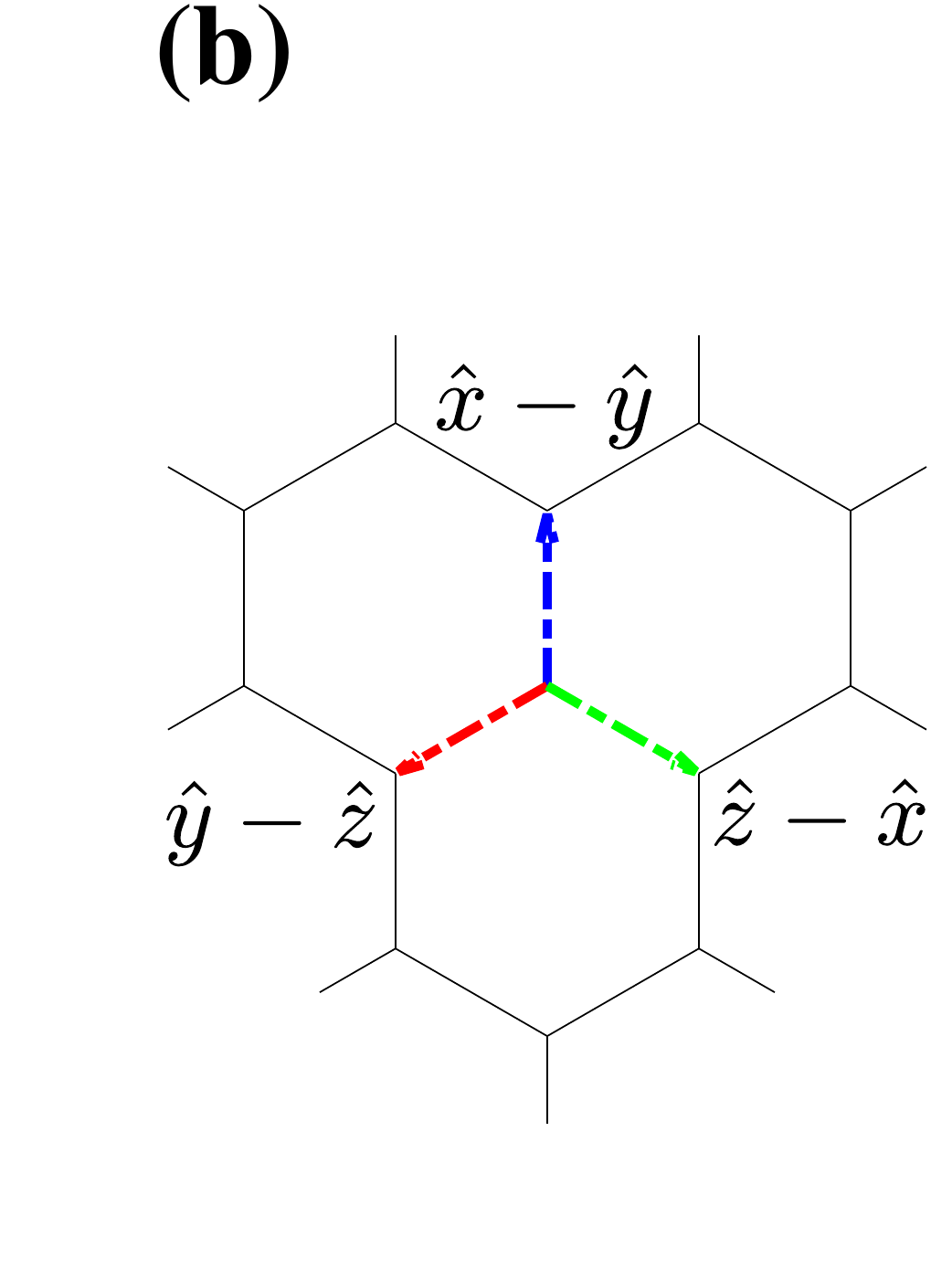}
\caption{(a) Field directions in the honeycomb system represented on the 
surface of a sphere. The dark plane is the lattice $ab$ plane. $\hat x$, 
$\hat y$, and $\hat z$ are the Kitaev axes of spins residing on the three 
different bonds [panel (b)]. Red, green, and blue planes are those normal to 
the respective spin axes. (b) Relation between lattice and spin basis vectors.}
\label{fig:Orientations}    
\end{figure}

In this Letter we analyze the low-energy physics of the field-induced 
magnetically disordered phase in $\alpha$-RuCl$_3$. We employ a slave-fermion 
representation to demonstrate that the properties of this state depend 
strongly on the direction of the applied field, which can produce a gapless 
Dirac QSL, a gapped chiral QSL, or a gapped and topologically trivial 
paramagnetic phase. By Variational Monte Carlo (VMC) calculations using 
Gutzwiller-projected states, we obtain a semi-quantitative description of 
the suppression of low-field order and deduce the dispersions and phase 
diagrams in all three cases. 

\begin{figure*}[t]
\includegraphics[width=6.4cm]{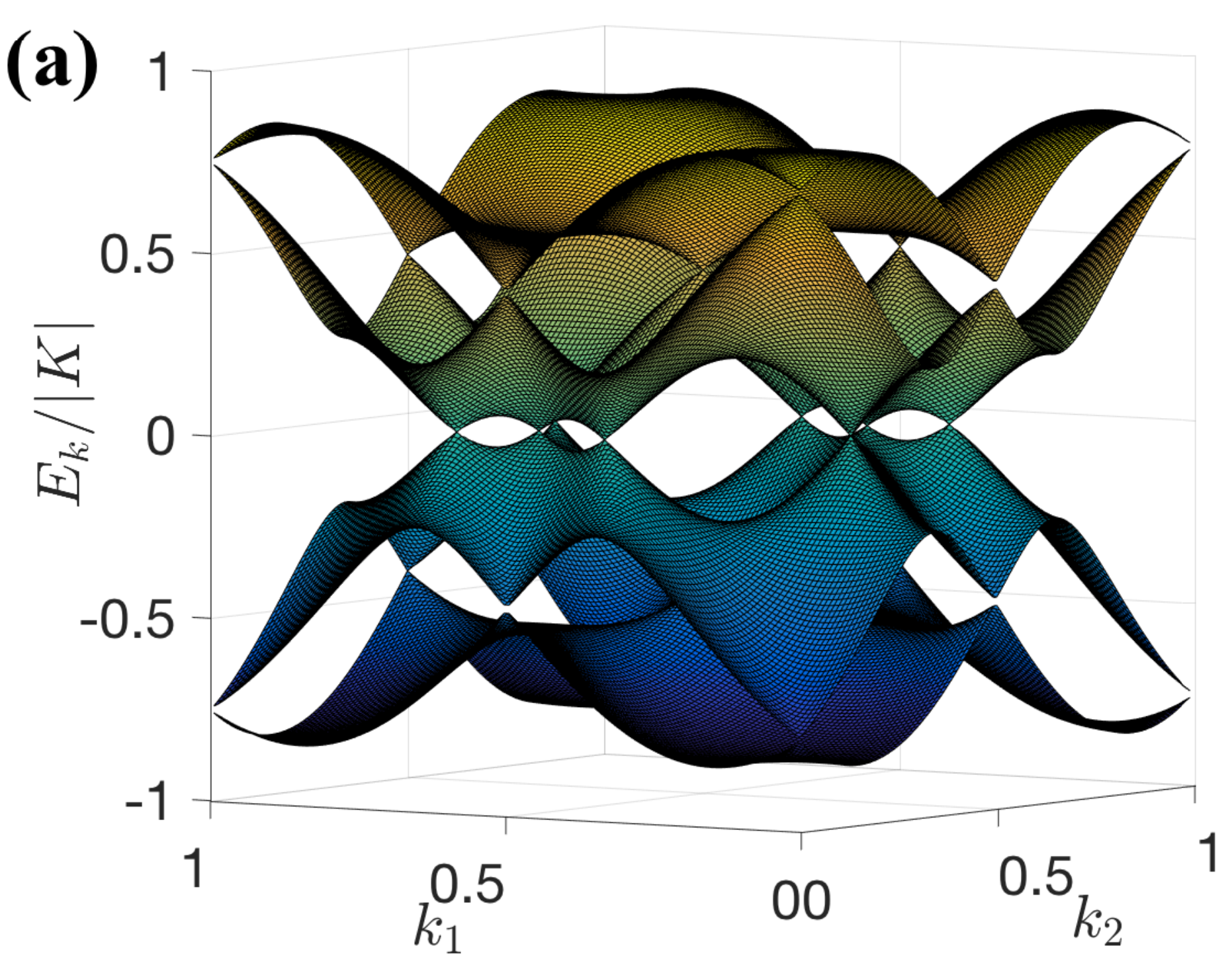}
\includegraphics[width=4.8cm]{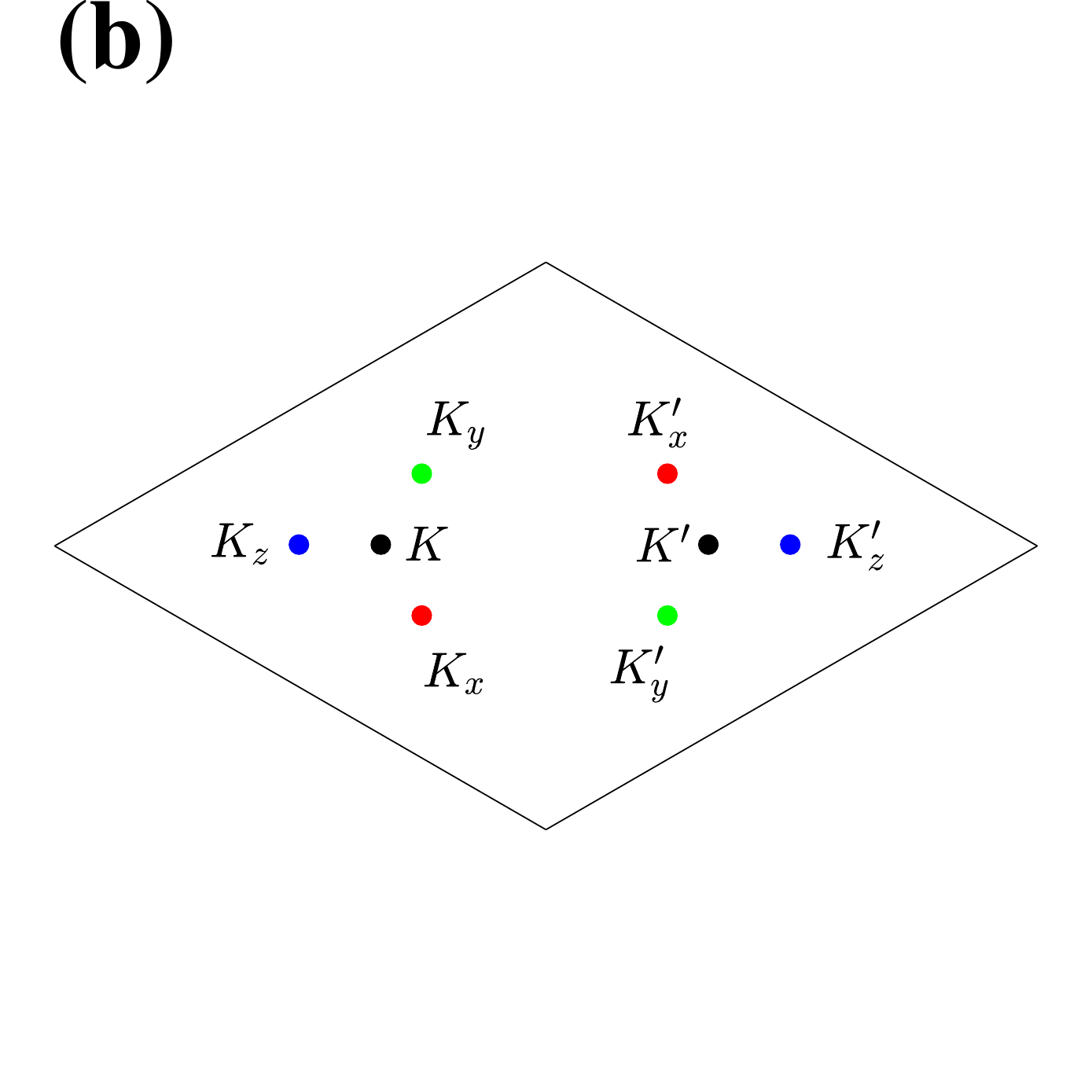}
\includegraphics[width=6.4cm]{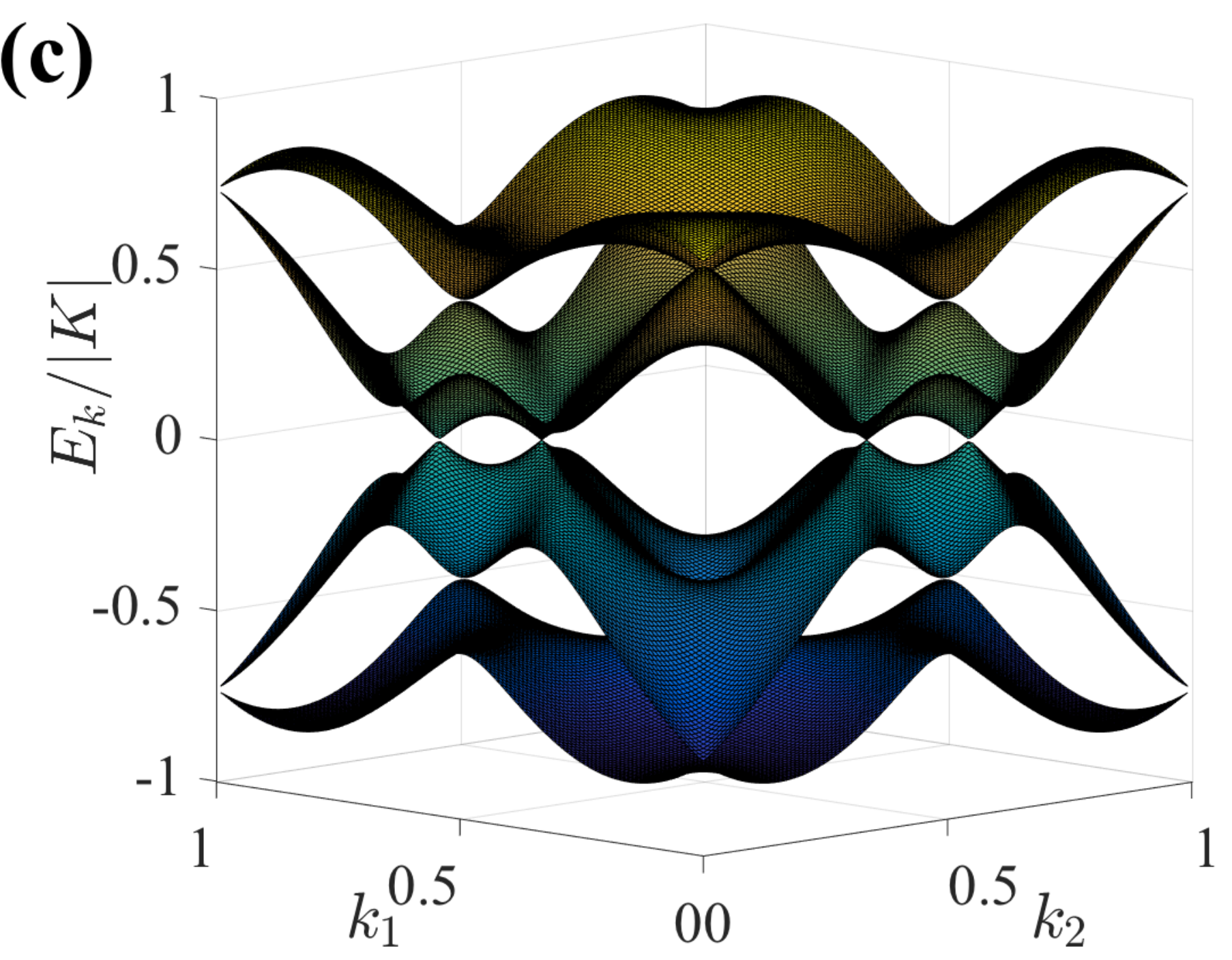}
\caption{(a) Spinon dispersion with $\pmb B = 0$, showing eight Dirac cones.
(b) Locations of the eight nodes in the Brillouin zone. (c) Spinon dispersion 
with an in-plane field $\pmb B \parallel ({\hat x} - {\hat y})$, showing two 
remaining pairs of Dirac cones close to $K, K'$ and $K_z, K_z'$.} 
\label{fig:Diraccones}    
\end{figure*}
      
We begin with a minimal effective model containing only Kitaev ($K$) and 
symmetric off-diagonal ($\Gamma$) terms, 
\begin{equation} \label{KGammaB}
H = \!\!\!\!\!\! \sum_{\langle i,j \rangle \in \alpha\beta(\gamma)} \!\!\! [K S_i^\gamma 
S_j^\gamma + \Gamma (S_i^\alpha S_j^\beta + S_i^\beta S_j^\alpha)] + g \mu_B \! 
\sum_i \! \pmb B \! \cdot \! \pmb S_i.
\end{equation}
We neglect Heisenberg exchange terms, which are argued to be small in a 
perturbative expansion \cite{KGammma_Li_arX16}. We adopt the representative 
parameters $K = - 6.8$meV and $\Gamma = 9.5$meV used to fit the spin-wave 
spectrum of $\alpha$-RuCl$_3$ measured by INS \cite{Ru_neutron_Wen_L17}, 
which places the system far from the (generic) Kitaev regime 
\cite{GaplessRespons_Balents_L16}. Within this model we treat 
the Land\'e $g$ factor as isotropic ($g = 2$) \cite{rjav}. 

Chemical coordination and spin-orbit coupling (SOC) in Na$_2$IrO$_3$ and 
$\alpha$-RuCl$_3$ fix the spin axes of the Kitaev interaction terms to the 
lattice such that the crystalline $c$ axis is the [111] direction in the spin 
frame, as represented in Fig.~\ref{fig:Orientations}(a). The lattice bond 
directions, which lie in the $ab$ plane and will turn out to have particular 
significance, are ($\hat x - \hat y$), ($\hat y - \hat z$), and ($\hat z -
\hat x$) in the spin basis [Fig.~\ref{fig:Orientations}(b)]. $\alpha$-RuCl$_3$ 
has space-group symmetry P$3_112$, whose point group is $D_3$. However, we 
restrict our considerations to a single honeycomb layer, which has point 
group $D_{3d}$ when the applied field $\pmb B = 0$, and any lattice rotation 
is associated with a spin rotation due to SOC. A QSL state should break 
neither translation nor any of the symmetries remaining in the presence 
of $\pmb B$.

Our analysis is based on a slave-particle representation in which spin 
operators at site $i$ are represented by two species of fermionic spinon, 
taking the quadratic forms $S_i^m = {\textstyle \frac12} C_i^\dag \sigma_m 
C_i$, where $C_i^\dag = (c_{i\up}^\dag \;\, c_{i\dn}^\dag)$, $m = x$, $y$, $z$, 
$\sigma_m$ are the Pauli matrices, and the spinons obey the on-site 
particle-number constraint $\hat N_i = c_{i\up}^\dag c_{i\up} + c_{i\dn}^\dag 
c_{i\dn} = 1$. By applying the mean-field approximation detailed in Sec.~S1 
of the Supplemental Material (SM) \cite{sm}, we express the $K$-$\Gamma$ 
model in the form 
\beq\label{Hmf}
H_{\rm mf} & = & \sum_{\langle ij \rangle \in \alpha\beta(\gamma)} [C_i^\dag (t_1^\gamma 
R_{\alpha\beta} - it_0^\gamma + t_2^\gamma \sigma_\gamma) C_j + {\rm h.c.}] \nonumber 
\\& & \;\;\;\;\;\;\;\; + g \mu_B \sum_i C_i^\dag ({\textstyle \frac12} \pmb B \! 
\cdot \! {\pmb \sigma} + \lambda) C_i + H_0,
\eeq
where $R_{\alpha\beta} = {-i\over\sqrt2}(\sigma_\alpha + \sigma_\beta)$, $t_1^\gamma
 = - {\textstyle \frac12} |K| \langle C_i^\dag R_{\alpha\beta} C_j \rangle^*$, 
$t_{0,2}^\gamma = - {\textstyle \frac{1}{8}}(\Gamma - |K|) [\langle C_i^\dag 
\sigma_\alpha R_{\alpha\beta} C_j \rangle^* \pm \langle C_i^\dag \sigma_\beta 
R_{\alpha\beta} C_j \rangle^*]$, $\lambda$, a Lagrange multiplier corresponding 
to the average particle-number constraint, functions as a chemical potential, 
and $H_0$ is a constant. The $t_1^\gamma$ and $t_2^\gamma$ terms are analogous to 
the Rashba SOC of electrons \cite{Rashba_Morais_B10}. We will determine all of 
these parameters by VMC calculations in which the local constraint is enforced 
exactly. 

Before turning to this quantitative treatment, we consider the qualitative 
nature of the mean-field spinon state and its response to a magnetic field. 
The mean-field state is by construction a QSL, which in Eq.~(\ref{Hmf}) has 
U(1) gauge symmetry because the spinon number is conserved. Finite spinon 
pairing terms may result in a Z$_2$ QSL \cite{You_Z2PSG_B12}, but are 
neglected in Eq.~(\ref{Hmf}) because we have found (Sec.~S1 of the SM 
\cite{sm}).that they are not favored energetically at intermediate magnetic 
fields.

The honeycomb lattice usually supports a conical spectrum due to its $C_{3v}$ 
point-group symmetry. In graphene the $K$ and $K'$ points are invariant under 
$C_{3v}$, whose two-dimensional irreducible representation results in two-fold 
energetic degeneracies and hence in Dirac cones. These cones can be gapped in 
two ways, one being to add a sublattice chemical potential, $\mu_z$, which 
breaks the symmetry down to $C_3$. The alternative, which does not break the 
symmetry explicitly, is that increasing strain causes the two cones to move 
together adiabatically until they merge into a single branch, whose 
``semi-Dirac'' \cite{semiDirac} dispersion is quadratic in one ${\bm k}$-space 
direction but remains linear in the other, after which a full gap opens.

For illustration, we take $t_1^\gamma = 1$, $t_0^\gamma = t_2^\gamma = 0$ 
in Eq.~(\ref{Hmf}). When $\pmb B = 0$, the spinon dispersion 
[Fig.~\ref{fig:Diraccones}(a)] is gapless with eight Dirac cones in 
the Brillouin zone [Fig.~\ref{fig:Diraccones}(b)]. These conical 
dispersions are protected by (emergent) symmetries, which separate 
them into two types. 

\noindent
(1) At $K$ and $K'$, $C_{3v}$ is preserved even with SOC and protects the 
cones. The generators of the $C_{3v}$ group are 
\beq
C_3 = e^{\mp i{2\pi\over3}{\mu_z\over2}} \! \otimes \! e^{-i {2\pi\over3}{\sigma_c\over2
\sqrt3}}, M_z = \mu_{\pm} \! \otimes \! {\textstyle \frac{1}{\sqrt2}} 
(\sigma_x \! - \! \sigma_y), \label{C3} 
\eeq
at $K$ and $K'$ respectively, where $\mu_m$ are Pauli matrices operating on 
the sublattice indices, $\mu_\pm = \pm {\textstyle \frac12} \mu_x + {\textstyle 
\frac{\sqrt3}{2}} \mu_y$, and $\sigma_c = \pmb \sigma \! \cdot \! {\hat c}$ 
with ${\hat c} = {1\over\sqrt3} (\hat x + \hat y + \hat z)$. Like a $\mu_z$ 
term, a field (Zeeman) term, $H' = B_c \sigma_c$, also breaks the mirror 
symmetry, gapping the cones at $K$ and $K'$. Unlike $\mu_z$, which creates 
a trivial gapped phase, $B_c$ gives a nonzero Chern number, as detailed in 
Sec.~S2 of the SM \cite{sm}. 

\begin{table}[b]
\centering
\begin{tabular}{c||c|c|c}
Dirac node & $\ $Mass$\ $ & Sign of mass & Chern number \\
\hline
$K,K'$    & $\pmb B \! \cdot \! \hat c$ & $-$ & $-$sgn($\pmb B \! \cdot \! 
\hat c$) \\
$K_x,K_x'$ & $\pmb B \! \cdot \! \hat x$ & $+$ & sgn($ \pmb B \! \cdot \! 
\hat x$)\\
$K_y,K_y'$ & $\pmb B \! \cdot \! \hat y$ & $+$ & sgn($ \pmb B \! \cdot \! 
\hat y$)\\
$K_z,K_z'$ & $\pmb B \! \cdot \! \hat z$ & $+$ & sgn($ \pmb B \! \cdot \! 
\hat z$)\\
\end{tabular}
\caption{Action of magnetic fields in giving mass to the Dirac cones of 
Fig.~\protect{\ref{fig:Diraccones}}(a).}\label{tab:mass}
\end{table}

\noindent
(2) At the $K_x$ and $K_x'$ points, expansion of Eq.~(\ref{Hmf}) in momentum 
space gives an effective Hamiltonian with $C_{4v}$ symmetry, whose generators 
are represented as 
\beq
C_4 =  e^{\mp i{\mu_z}{\pi\over4}} \otimes e^{i{\sigma_x}{\pi\over4}},\ M_x = \mu_x 
\otimes {\textstyle \frac{1}{\sqrt2}} (\sigma_y - \sigma_z), \label{C4}
\eeq 
at $K_x$ and $K_x'$ respectively. Because the momentum itself is not invariant 
under $C_{4v}$, this is an emergent symmetry, which is interpreted as operations 
with sublattice-spin coupling but without spatial rotation. As above, a 
magnetic field along $\hat x$ gaps the cones, with the key difference that 
the mass $B_x$ has the opposite sign to the mass $B_c$ for the cones at $K$ 
and $K'$ [{\it cf}.~Eqs.~(\ref{C3}) and (\ref{C4})], which has important 
consequences for the total Chern number. The same physics applies to the 
Dirac cones at $K_y, K_y'$ and $K_z, K_z'$ in fields $B_y$ and $B_z$. 

These results are summarized in Table \ref{tab:mass}, where we have assumed 
that the Dirac cones are independent. Clearly if the magnetic field is oriented 
such that $\pmb B \perp \hat \alpha$ and $\pmb B \perp \hat \beta$ ($\alpha, 
\beta = c, x, y, z$), i.e.~at any intersection of the circles on the sphere 
in Fig.~\ref{fig:Orientations}(a), then two pairs of Dirac cones have 
vanishing masses, as shown in Fig.~\ref{fig:Diraccones}(c). These cones 
are symmetry-protected and cannot be gapped individually. With increasing 
field, the cones of each pair move together, merging to a semi-Dirac 
dispersion, and then open a full gap, as shown in Sec.~S3 of the SM \cite{sm}. 

For fields $\pmb B\perp \hat \alpha$, i.e.~any other points on the circles 
in Fig.~\ref{fig:Orientations}(a), one pair of Dirac cones retains a 
vanishing linear mass and one may expect the spinons to remain gapless. 
However, higher-order processes generate a small gap that scales algebraically 
with the field, $\Delta =  c|\pmb B|^a$. We present some numerical results 
for $a$ in Sec.~S3 of the SM \cite{sm}; as an example, $a = 2.4$ if $\pmb 
B \parallel (\hat x + \hat y)$. The resulting slow growth of this gap may 
cause it to remain below the measurement temperature over a broad field 
range, making the system behave as if it were still gapless. This  
explains qualitatively the appearance of point-node excitations for all 
in-plane field directions in NMR \cite{Ru_FieldNMR_arX17}.

For all other field directions, all the cones have nonzero masses and the 
spinon dispersion has a gap, which opens linearly in $|\pmb B|$. The total 
Chern number of the half-filled fully gapped bands is $\mathcal C = {\rm sgn} 
(\pmb B \! \cdot \! \hat x) + {\rm sgn}(\pmb B \! \cdot \! \hat y) + {\rm 
sgn}(\pmb B \! \cdot \! \hat z) - {\rm sgn}(\pmb B \! \cdot \! \hat c)$,
which is either 0 or $\pm 2$. If a point on the sphere of 
Fig.~\ref{fig:Orientations}(a) is surrounded by only three arcs, then 
$\mathcal C = \pm 2$ and a chiral QSL is obtained, which persists up to a 
critical value of $|\pmb B|$ where its gap closes (Sec.~S3 of the SM 
\cite{sm}). If a point is surrounded by four arcs, then $\mathcal C = 0$ 
and the gap never closes at finite $|\pmb B|$, meaning that the system is 
connected adiabatically to the fully polarized trivial (direct-product) 
state.

The topological transition between the $\mathcal C = 0$ and $\pm 2$ 
regimes is a function of field angle and is discussed in Sec.~S4 of the SM 
\cite{sm}. We recall that the field-induced QSL states we find have U(1) 
gauge symmetry, in contrast to the Z$_2$ gauge symmetry and finite 
vison gap of the Kitaev QSLs. In two spatial dimensions, U(1) gauge fields 
are confined, which is the trivial gapped phase, unless the matter (spinon) 
field is gapless, which is our four-cone case, or there is a Chern-Simons 
term, which is our $\mathcal C = \pm 2$ case. The transition may therefore be 
considered as a spinon (de)confinement process and its position obtained 
from the spinon dispersion and Chern number.

\begin{figure}[t]
\centering
\includegraphics[width=8.0cm]{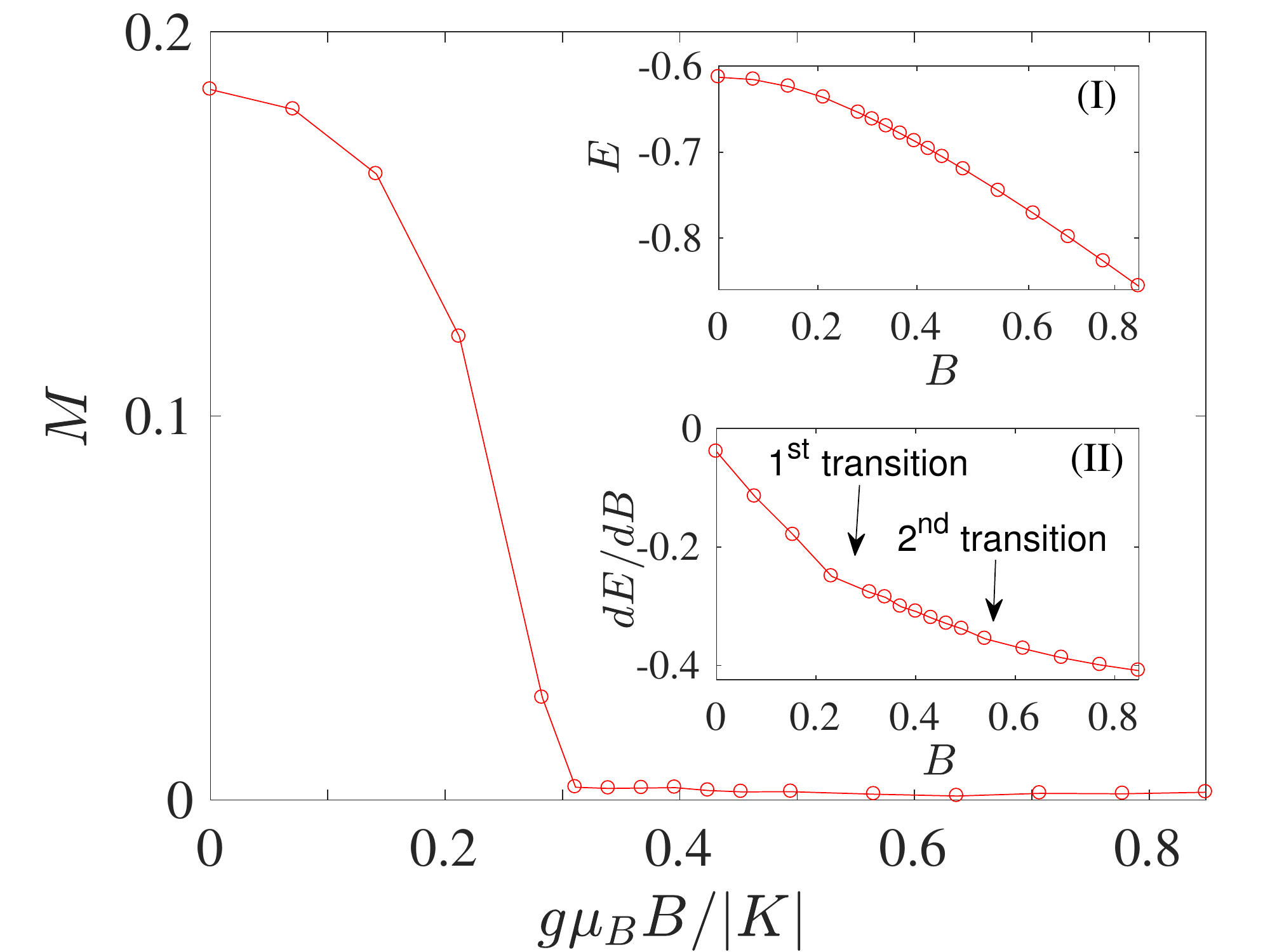}
\caption{Magnetization as a function of field for $\pmb B \parallel (\hat x
 - \hat y)$. Inset (I) shows the ground-state energy, $E$, as a function of 
$B$. Inset (II) shows $dE/dB$, whose discontinuity coincides with the 
magnetic transition. At the weak second transition, the Dirac cones become 
gapped.}
\label{fig:M}
\end{figure}

Thus at the mean-field level we have obtained crucial qualitative insight 
into the physical properties of three different field-induced disordered 
phases. However, the ground state observed in $\alpha$-RuCl$_3$ is a zigzag 
magnetic order \cite{Fletcher_JCSA_1967, Sears_PRB_91, Johnson_PRB_92, 
Cao_PRB_93}, which some authors \cite{KJGamma_L14} find in the classical 
$K$-$\Gamma$ model (\ref{KGammaB}), whereas others \cite{rjav} find that a 
small ferromagnetic (FM) Heisenberg interaction is required to stabilize it. 
In Sec.~S1 of the SM \cite{sm} we discuss the quantum model and show that our 
results are reinforced by FM terms. To compare with experimental data, we 
introduce zigzag order within the single-$Q$ approximation, 
$\pmb M_i = M (\sin\eta [\hat e^x_i \cos (\pmb Q \! \cdot \! \pmb r_i) + 
\hat e^y_i \sin (\pmb Q \! \cdot \! \pmb r_i)] + \cos \eta \hat e^z_i)$, 
where $\pmb Q = [1/2,1/2]$, the local (spin) axes, $\hat e_i^\alpha$, are 
fixed by the classical ground state, and $\eta$ is the canting angle. We 
treat the static order as an external field to obtain a new mean-field 
Hamiltonian, $H_{\rm mf}' = H_{\rm mf} - \sum_{i} ({\textstyle \frac12} \pmb 
M_i\! \cdot \! C_i^\dag \pmb \sigma C_i + {\rm h.c.})$; in a variational 
treatment, this process is equivalent to introducing an additional 
decoupling channel in $H_{\rm mf}$ (Sec.~S1 of the SM \cite{sm}). 

VMC calculations are based on the mean-field states but enforce the local 
constraint on spinon number by Gutzwiller projection. This method has 
been applied widely to capture the essential physics of correlated electron 
systems and in certain cases, including slave-parton approaches to QSL 
states, can provide exact information. Here we employ the variational wave 
functions $|\psi_{\rm G}(\pmb p) \rangle = P_G| \psi_{\rm mf} (\pmb p) \rangle$, 
where $\pmb p$ denotes ($t_0^\gamma,t_1^\gamma,t_2^\gamma,\lambda,M,\eta$), to 
obtain the optimal state by minimizing the ground-state energy, $E = {\langle 
\psi_{\rm G} |H| \psi_{\rm G} \rangle \over \langle \psi_{\rm G}| \psi_{\rm G} 
\rangle}$, on a system of 128 sites and use it to compute physical expectation 
values. To benchmark the accuracy of our results, in a small (8-site) system we 
obtain an overlap $\langle \psi_{\rm VMC} | \psi_{\rm ED} \rangle = 0.988$ between 
the VMC and exact-diagonalization wave functions (more details are provided 
in Sec.~S1 of the SM \cite{sm}). 

We establish magnetic phase diagrams by fixing the field direction and 
increasing its magnitude. Zigzag order is suppressed for all field 
directions and vanishes at a lower critical field (Fig.~\ref{fig:M}), 
beyond which, as anticipated from the mean-field analysis, three different 
field-induced (partially polarized) disordered phases appear. 

\begin{figure}[t]
\centering
\includegraphics[width=8.3cm]{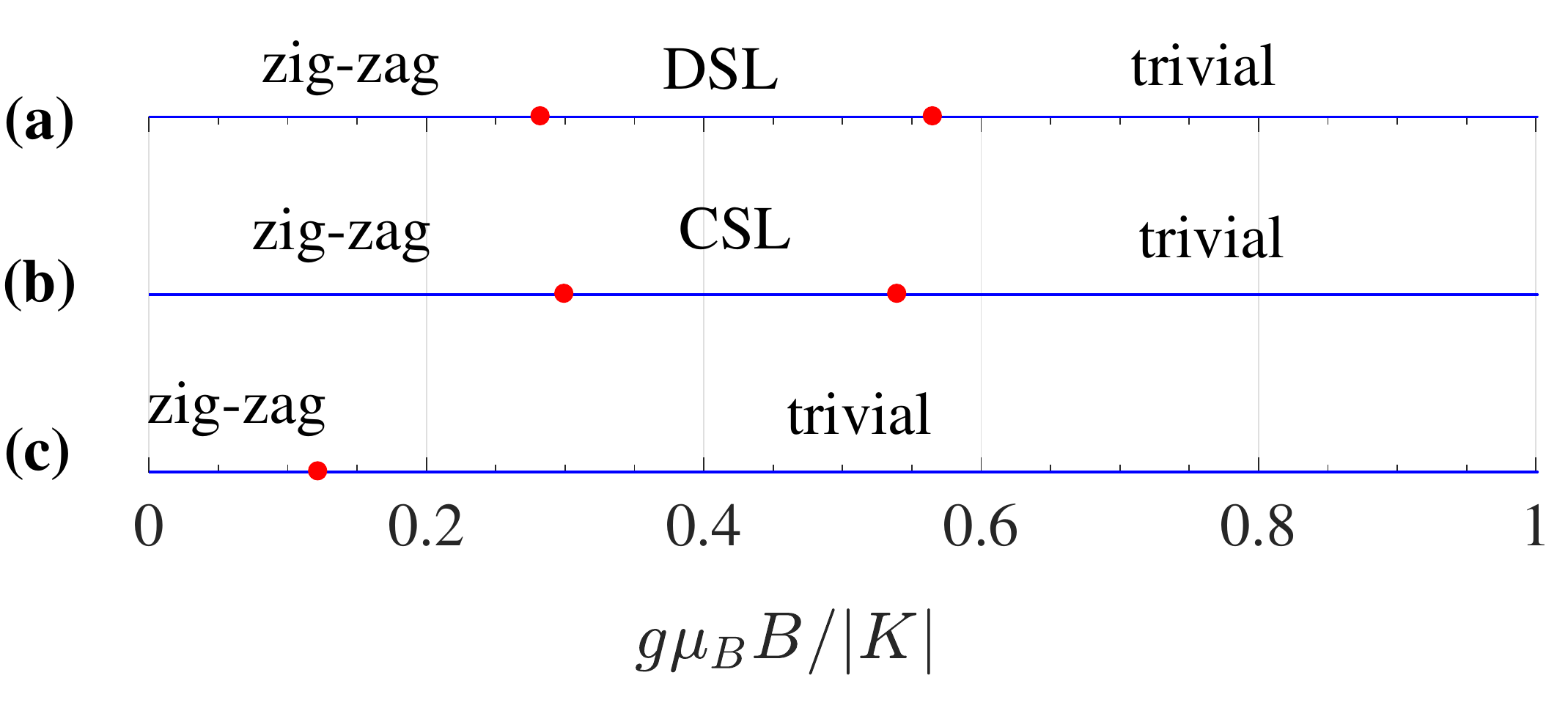}
\caption{Phase diagrams for different orientations of $\pmb B$. (a) $\pmb 
B \parallel (\hat x - \hat y)$, where the intermediate phase is a field-induced 
Dirac spin liquid (DSL). Both transitions are of second order. (b) $\pmb B 
\parallel (\hat x + \hat y + {1\over 2} \hat z)$, where the intermediate 
phase is a field-induced chiral spin liquid (CSL). The lower transition is of 
first order, while the upper is second-order. (c) $\pmb B \parallel (\hat x
 + \hat y - 2\hat z)$, where there is only one, weakly first-order, 
transition.}    
\label{fig:pd}    
\end{figure}

\noindent
{\it Dirac QSL}. If $\pmb B \parallel \hat \alpha$ or $\pmb B \parallel 
(\hat \alpha - \hat \beta)$ ($\alpha,\beta = x,y,z$), there are two critical 
points, as shown in Fig.~\ref{fig:pd}(a). The first marks the continuous 
transition from the ordered phase to a disordered one (Fig.~\ref{fig:M}) 
in which the spinon dispersion, with optimized parameters determined from 
VMC, is gapless. Thus the intermediate phase is a stable U(1) Dirac QSL. At 
the second transition, which is also continuous, a gap opens as the system 
enters the trivial paramagnetic phase. 

\noindent
{\it Chiral QSL.} If $\mathcal C = \pm 2$, for example when $\pmb B 
\parallel \hat c$ or $\pmb B \parallel (\hat x + \hat y + {1\over 2} 
\hat z)$ [Fig.~\ref{fig:pd}(b)], the disordered phase is a chiral QSL
\cite{KalmayerLaughlin, WenWilczekZee}. This state is gapped, Abelian, has 
chiral edge modes, supports semionic spinon excitations, and has an integer-quantized 
thermal Hall conductivity, which can be measured in experiment. It exists 
over a continuous regime of applied field directions, upon which the 
critical field depends strongly (numerically determined values include $g 
\mu_B B/|K| \simeq 2.4$ for $\pmb B \parallel \hat c$ and 0.3 for $\pmb B 
\parallel (\hat x + \hat y + {1\over 2} \hat z)$ [Fig.~\ref{fig:pd}(b)]). 
At a higher critical point, the chiral QSL undergoes a transition to the 
trivial phase.

\noindent
{\it Gapped paramagnet.} For field orientations giving $\mathcal C = 0$, 
and for the cases where the field direction lies on the circles in 
Fig.~\ref{fig:Orientations}(a), the phase diagram has only one critical 
point [Fig.~\ref{fig:pd}(c)]. This separates the ordered phase from the 
trivial polarized phase, whose gap opens linearly with field, except on 
the special lines where it opens algebraically with a higher power. 

We have analyzed the spin-liquid state within a slave-particle representation,
which is uncontrolled. However, VMC studies allowing accurate enforcement of 
the constraint reveal only moderate quantitative alterations to the results. 
This indicates that the spinon description is well able to capture all the 
significant magnetic degrees of freedom. We stress that the gapless spin 
excitations of the field-induced U(1) QSL are deconfined Dirac fermions, 
which are different from the Majorana fermions in the gapless Kitaev QSL. 
Indeed, our results show that Kitaev QSLs are poor trial states, giving much 
higher energies than the U(1) QSL for the model of Eq.~(\ref{KGammaB}) with 
$\Gamma/|K| = 1.4$. To address the stability of the U(1) QSL, we have tested 
the possibility of spinon pairing, but show in Sec.~S1 of the SM \cite{sm} 
that this is not favorable. However, here we do not try to classify and test 
all possible QSL states of the model. We comment that recent numerical studies 
of the $K$-$\Gamma$ model \cite{Catuneanu_arX17,KGammma_DMRG_YBKim_arX17} also 
find QSL states over much of the phase diagram, albeit without detailed 
consideration of magnetic order or applied fields.

Our conclusions are in quantitative agreement with experiments on 
$\alpha$-RuCl$_3$. The lower critical fields for the loss of zigzag order 
fall around 7.5 T for in-plane ($ab$) fields. Our VMC phase diagrams in this 
case show that the field-induced disordered phase is either truly gapless for 
discrete field directions or otherwise has a very small gap that grows 
algebraically with $|\pmb B|$, such that for temperatures $T \ge 1.5$ K it 
would appear gapless. These results are fully consistent with recent NMR 
observations of a spin-lattice relaxation rate $1/T_1 \propto T^3$ and a 
largely isotropic response for all in-plane field orientations. 

In summary, we have studied the $K$-$\Gamma$ model on the honeycomb lattice 
with an external magnetic field. By using Gutzwiller-projected states as 
variational wave functions and including zigzag magnetic order, we find 
three different field-induced disordered phases, whose nature varies 
strongly with the field direction. In certain cases, the intermediate QSL 
is gapless with Dirac-cone excitations, which are protected by emergent 
symmetries. In others it is a gapped chiral QSL, which may be sought in 
experiment through its integer-quantized thermal Hall effect. 

{\it Acknowledgments.} We are grateful to W. Yu and J. Wen for the 
experimental collaboration which initiated this study. We thank W. Ku, 
J.-X. Li, H.-J. Liao, C. Morais Smith, R. Valent\'{\i}, M. Vojta, F. Wang, 
X.-Q. Wang, F. Wilczek, T. Xiang, J.-Z. Zhao, and Y. Zhou for helpful 
discussions. This work was supported by the NSF of China (Grant No.~11574392), 
the Ministry of Science and Technology of China (Grant No.~2016YFA0300504), 
and the Fundamental Research Funds for the Central Universities and the 
Research Funds of Renmin University of China (No.~15XNLF19). 



\clearpage

\setcounter{equation}{0}
\renewcommand{\theequation}{S\arabic{equation}}
\setcounter{figure}{0}
\renewcommand{\thefigure}{S\arabic{figure}}
\setcounter{section}{0}
\renewcommand{\thesection}{S\arabic{section}}
\setcounter{table}{0}
\renewcommand{\thetable}{S\arabic{table}}

\onecolumngrid

\vskip1cm

\centerline{\large {\bf {Supplemental Material for ``Dirac and chiral quantum 
spin liquids}}}

\vskip1mm

\centerline{\large {\bf {on the honeycomb lattice in a magnetic field''}}}

\vskip4mm

\centerline{Zheng-Xin Liu and B. Normand}

\vskip8mm

\twocolumngrid

\section{Slave-particle approach for the quantum spin liquid}\label{Fermion}

\subsection{Fermionic representation of the $K$-$\Gamma$ interaction}
\label{FermionA}

In the slave-particle representation employed here, the spin operators at 
every site $i$ are represented by two fermions, $c_{i\up}^\dag$ and $c_{i\dn}^\dag$. 
Because we are concerned only with the spin degrees of freedom, we use the 
terms ``slave fermion'' and ``spinon'' interchangeably. As stated in the main 
text, the spin operators are expressed as 
\beq\label{SpinO}
S_i^m = {\textstyle \frac12} C_i^\dag \sigma_m C_i,\ \ \ m = x, y, z,
\eeq
where $C_i^\dag = (c_{i\up}^\dag \,\, c_{i\dn}^\dag)$ and $\sigma_m$ are the Pauli 
matrices. The space of physical spin states is spanned by the sector with 
only one spinon per site, giving a local constraint on the net particle 
number, $\hat N_i = c_{i\up}^\dag c_{i\up} + c_{i\dn}^\dag c_{i\dn} = 1$. It 
has been shown that this representation has an SU(2) gauge symmetry 
\cite{SU(2)gauge}. In terms of these fermionic operators, the conventional 
two-spin interactions are given up to some unimportant constants by 
\beq
\pmb {S}_{i} \! \cdot \! \pmb {S}_{j} = - {\textstyle \frac{1}{4}} (C_i^\dag C_j 
C_j^\dag C_i + C_i^\dag \bar C_j\bar C_j^\dag C_i) \label{singlet_DC}
\eeq
for the antiferromagnetic Heisenberg exchange interaction, where 
${\bar C}^\dag = (- c_\dn \,\, c_\up)$, and 
\beq
S^m_i S^m_j & = & - {\textstyle \frac{1}{8}} (C_i^\dag C_j C_j^\dag C_i + C_i^\dag 
\bar C_j \bar C_j^\dag C_i \\ & & \;\;\;\; + C_i^\dag \sigma_m C_j C_j^\dag 
\sigma_m C_i + C_i^\dag \sigma_m \bar C_j\bar C_j^\dag \sigma_m C_i) \nonumber
\label{Ising}
\eeq
for the Ising interaction.

To express the $K$-$\Gamma$ interactions in fermionic operators, it is 
advantageous to perform a local basis rotation on one site ($j$) of each 
bond. For the example of the $z$-bond, the spin axes are rotated by $\pi$ 
around the direction $\hat x + \hat y$, giving 
$${S_j^m}' = {\textstyle \frac12} C_j'^{\dag} {\sigma_m} C_j'$$ 
with $C_j' = R_{xy} C_j$ and ${\bar C}_j' = R_{xy} {\bar C}_j$, where  
$$R_{xy} = e^{-i {\pi\over 2\sqrt2} (\sigma_x +\sigma_y)} = - {\textstyle \frac{i}{\sqrt2}} 
(\sigma_x + \sigma_y).$$ The result is a transformation of the spin operators 
to $$S_j^z = - {S_j^z}', \; S_j^x = {S_j^y}', \; S_j^y = {S_j^x}'.$$ Now the 
$z$-bond interaction term, 
\beq\label{KGz}
K S^z_i S^z_j  + \Gamma (S_i^x S_j^y + S_i^yS_j^x),
\eeq
takes the form of an XXZ interaction of the new spin operators, 
\Beq
|K| & & S^z_i S_j^{z'}  + \Gamma (S_i^x S_j^{x'} + S_i^y S_j^{y'}) \\ & & = |K| \, 
\pmb S_i \! \cdot \! \pmb S_j' + (\Gamma - |K|)[S_i^x S_j^{x'} + S_i^y S_j^{y'}], 
\Eeq
which in fermionic form is 
\Beq
& & - {\textstyle \frac{1}{4}} \Gamma (C_i^\dag C_j' C_j'^\dag C_i + C_i^\dag 
\bar C_j' \bar C_j'^\dag C_i) \\ & & -  {\textstyle \frac{1}{8}} 
(\Gamma - |K|) [ C_i^\dag \sigma_x C_j' C_j'^\dag \sigma_x C_i + C_i^\dag \sigma_x 
\bar C_j' \bar C_j'^\dag \sigma_x C_i \\ & & \ \ \ \ \ \ \ \ \ \ \ \ \ \ \ \ 
+ C_i^\dag \sigma_y C_j' C_j'^\dag \sigma_y C_i + C_i^\dag \sigma_y \bar C_j' 
\bar C_j'^\dag \sigma_y C_i].
\Eeq
This procedure is repeated for the interactions on the $x$- and $y$-bonds. 

\subsection{Mean-field decoupling and spin-liquid symmetry}
\label{sec:u1orz2}

By gathering all terms and applying the mean-field approximation $- O^\dag O 
\approx - \langle O^\dag \rangle O - \langle O \rangle O^\dag + \langle O^\dag 
\rangle \langle O \rangle$, where our choice of sign specifies attractive 
interaction terms \cite{Nagaosabook}, we decouple the Hamiltonian to obtain 
a quadratic description of the quantum spin-liquid (QSL) state. In the complete 
fermionic representation of the $K$-$\Gamma$ Hamiltonian, both spinon hopping 
and spinon pairing terms should have finite expectation values and the full 
mean-field Hamiltonian is 
\beq\label{Hmf}
H_{\rm mf} & = & \!\!\!\!\! \sum_{\langle ij \rangle \in \alpha\beta(\gamma)} \!\!\!\! 
\{[C_i^\dag (t_1^{\gamma} R_{\alpha\beta} \! - \! it_0^{\gamma} \! + \! t_2^{\gamma} 
\sigma_\gamma \! + \! t_3^\gamma \sigma_\gamma R_{\alpha\beta}) C_j \nonumber \\ 
& &  + C_i^\dag (\Delta_1^{\gamma} R_{\alpha\beta} \! - \! i \Delta_0^{\gamma} \!
 + \! \Delta_2^{\gamma} \sigma_\gamma \! + \! \Delta_3^\gamma  \sigma_\gamma 
R_{\alpha\beta}) \bar C_j] + {\rm h.c.}\} \nonumber \\ & & + \sum_i C_i^\dag 
[{\textstyle \frac12} (g \mu_B \pmb B - \pmb M_i) \! \cdot \! {\pmb \sigma}
 + \lambda_i) C_i + H_0,
\eeq
where $H_0$ is an irrelevant constant. $\pmb M_i$ allows magnetic decouplings 
of the spinon Hamiltonian and is set to zero in Eq.~(2) of the main text for 
the purposes of analyzing the unrestricted mean-field solutions; its role is 
restored on P3 of the main text for the VMC analysis. $\lambda_i$ is a 
Lagrange multiplier for the particle-number constraint on each site, but 
by translational invariance all $\lambda_i$ have the same value, $\lambda$, 
which functions as the chemical potential (ensuring that the system remains 
half-filled with spinons for any value of $\pmb B$). The mean-field 
spinon-pairing parameters, $\Delta_{0,1,2,3}^\gamma$, are off-diagonal analogs 
of the spinon-hopping parameters, $t_{0,1,2}^\gamma$, which are defined below 
Eq.~(2) of the main text, and $t_{3}^\gamma = - {\textstyle \frac{1}{8}}(|K|
 - \Gamma) \langle C_i^\dag \sigma_\gamma R_{\alpha\beta} C_j \rangle^*$. While 
the parameters $t_1^\gamma$ and $\Delta_1^\gamma$ are finite for all finite 
$|K|$, $t_{0,2}^\gamma$ and $\Delta_{0,2}^\gamma$ are finite when $\Gamma > |K|$ 
and $t_{3}^\gamma$ and $\Delta_3^\gamma$ when $\Gamma < |K|$. Hence we ignore 
the last pair of parameters in the present study. 

In the absence of spinon pairing ($\Delta_{0,1,2}^\gamma = 0$), spinon number 
is conserved and the mean-field Hamiltonian (\ref{Hmf}) has a U(1) gauge 
symmetry, which is known as the invariant gauge group \cite{IGG}. If both 
the hopping and pairing parameters are nonzero, then the invariant gauge 
group is generally reduced to Z$_2$. The constants $t_{0,1,2}^\gamma$ and 
$\Delta_{0,1,2}^\gamma$ are treated as variational parameters, which are 
determined by minimizing the ground-state energy, either at the mean-field 
level, where the spinon-number constraint is enforced only globally, or by 
any more sophisticated technique. Here we use these constants as variational 
parameters in variational Monte Carlo (VMC) calculations, where the local 
constraint is enforced exactly and the values of the optimal parameters 
determine the (U(1), Z$_2$, or other) nature of the ground state. We defer 
the results of this analysis to the following subsection. 

\subsection{Variational Monte Carlo}\label{sec:VMC}

The essential physics of the VMC approach is that the local constraint is 
enforced by Gutzwiller projection. The variational parameters optimizing the 
projected state are determined by energy minimization (see main text) using 
Monte Carlo methods and all physical quantities can be calculated from this 
optimal state. Thus although VMC is based on the mean-field solution, it is 
far more advanced than the mean-field level, and as such has been used to 
gain insight into the physics of strongly interacting electron systems 
ranging from high-temperature superconductors to fractional quantum Hall 
liquids. In the context of QSLs, the Gutzwiller-projected state provided 
by VMC can be used to construct the exact ground state of certain exactly 
solvable models, including the pure Kitaev model on the honeycomb lattice 
\cite{Kitaev2006} and the Affleck-Kennedy-Lieb-Tasaki model on the spin-1 
chain \cite{AKLT}. Because the classification theory of many QSL states is 
based on slave-parton mean-field methods, VMC can provide the key information 
as to how the elementary excitations are fractionalized in these cases.

\subsubsection{Variational Hamiltonian}

We comment that there are two approaches to analyzing the competitiveness of 
a magnetically ordered state. One is to include a magnetic decoupling channel 
in the spin Hamiltonian, as shown in Eq.~(\ref{Hmf}).  The other is to include 
an effective external field that induces the magnetic order, as we do on P3 
of the main text to obtain the Hamiltonian $H'_{\rm mf}$. Because the amplitude, 
$|\pmb M|$, of the (zigzag) ordered component is a variational parameter, these 
two approaches are entirely equivalent in VMC calculations. In the present 
study, we have followed the second approach because our aim is to analyze a 
model reproducing the physics observed in experiment, without dwelling on the 
complexities encountered by other authors who have studied variations of the 
same problem (which we discuss in Sec.~\ref{s1d}). 

\begin{table}[b]
\centering
\begin{tabular}{c||c|c|c}
Energy & $\pmb B = 0.2 [1,-1,0]$ & $\pmb B = 0.4 [1,-1,0]$ & $\pmb B = 0.5 
[1,-1,0]$ \\ \hline
$E_{\rm KSL}$  & $- 0.6383$ & $- 0.7391$ & $- 0.7952$ \\
$E$          & $- 0.6549$ & $- 0.7455$ & $- 0.7990$ \\
$E_0$        & $- 0.6550$ & $- 0.7456$ &  $- 0.7991$ \\
\end{tabular}
\caption{Ground-state energies obtained from VMC calculations with 
different applied magnetic fields (taken for this comparison to have the 
same orientation). $E_{\rm KSL}$ is the energy of the optimized KSL state. 
$E$ denotes the energy obtained by allowing the variational parameters 
$\Delta_{0,1,2}^\gamma$ to change freely, $E_0$ the energy obtained by setting 
$\Delta_{0,1,2}^\gamma$ to zero. Fields and energies are quoted in units of 
$|K|$.}\label{tab:energy}
\end{table}

\subsubsection{Variational wave functions}

In the course of our variational analysis, we have tested the trial 
wavefunctions of U(1) QSL, Z$_2$ QSL, Kitaev-type spin-liquid (KSL, a Z$_2$ 
QSL state whose dispersion has two Majorana cones in the first Brillouin 
zone), and partially polarized zigzag-ordered states. The mean-field 
decoupling of the KSL may be found in Ref.~\cite{KGammma_DMRG_YBKim_arX17}. 
These studies ascertained that the optimal state we obtain is as close as 
possible to the true ground state. For the parameter regime of our study 
(Sec.~\ref{s1d}), which is that inspired by experiment, we find that the 
competition is always between zigzag order and U(1) QSL (of Dirac or chiral 
types) states. 


Regarding the competitiveness of the Z$_2$ wave functions at intermediate 
fields, we resume our discussion of spinon pairing terms within the VMC 
framework. When particle number is conserved in the mean-field Hamiltonian 
(i.e.~without spinon pairing), the projected state takes the form 
\beq\label{det}
\!\!\!\!\!\! |\psi_{\rm G}(\pmb p) \rangle_{U(1)} \! = \! P_G| \psi_{\rm mf} (\pmb 
p) \rangle \! = \! C_0 \! \sum_\alpha \! \det A(\pmb p,\alpha) |\alpha \rangle, 
\eeq
where $|\psi_{\rm mf}(\pmb p) \rangle$ is the mean-field ground state with 
variational parameters $\pmb p$, $|\alpha \rangle$ is the Ising basis, and 
$C_0$ is a normalization constant. $A(\pmb p, \alpha)$ is an $N$$\times$$N$ 
matrix with components $A_{jk} (\pmb p,\alpha) = \langle 0| c_{j,\alpha_j} 
\psi_k^\dag| 0 \rangle$, where $c^\dag_{j,\alpha_j}$ (Sec.~\ref{FermionA}) is 
the spinon creation operator at site $j$, with spin component $\alpha_j$, 
$\psi_k$ is the $k$th eigenmode of the mean-field Hamiltonian, which is 
occupied in the mean-field ground state, and $|0\rangle$ specifies the 
vacuum state. By contrast, in the presence of spinon-pairing terms, the 
ground state of the mean-field Hamiltonian is a BCS-type wave function,
$|\psi_{\rm BCS} (\pmb p) \rangle = \prod_{ij,\sigma\sigma'} [1 + a_{i\sigma,j\sigma'} 
(\pmb p) c_{i,\sigma}^\dag c_{j,\sigma'}^\dag] |0 \rangle$, where $a_{i\sigma,j\sigma'}$ 
is the wave function of two spinons in a Cooper pair. After Gutzwiller 
projection, the Z$_2$ spin-liquid state takes the form
\beq\label{pf}
\!\!\!\!\!\! |\psi_{\rm G} (\pmb p) \rangle_{Z_2} \! = \! P_G|\psi_{\rm BCS} 
(\pmb p) \rangle \! = \! C_0 \! \sum_\alpha \! {\rm Pf} B(\pmb p,\alpha) 
|\alpha \rangle, 
\eeq
where $B(\pmb p,\alpha)$ is an $N$$\times$$N$ skew-symmetric matrix with 
components $B_{ij}(\pmb p,\alpha) = a_{i\alpha_i,j\alpha_j}(\pmb p)$. 

We consider applied fields with different orientations and with a magnitude 
above the first critical field (main text), such that the low-field 
magnetic order is completely suppressed. The optimal wave functions given by 
our VMC calculations are such that all three spinon-pairing parameters are 
always very small, with ${\Delta_{0,1,2}^\gamma/ t_1^\gamma} \approx 10^{-2}$. 
Because of the BCS-type nature of the wave function of a Z$_2$ spin liquid 
[Eq.~(\ref{pf})], the pairing parameters, $\Delta_{0,1,2}^\gamma$, are never 
identically equal to zero in the variational process. Thus we repeat the 
calculation by fixing $\Delta_{0,1,2}^\gamma$ to zero. The results, displayed 
in Table \ref{tab:energy}, show clearly that the energy is unchanged or 
falls even lower, meaning that the U(1) spin liquid is favored energetically. 
For this reason we have neglected the spinon-pairing terms [second line of 
Eq.~(\ref{Hmf})] in all of our considerations in the main text. 

In Table \ref{tab:energy} we show also the energy of the optimal KSL state, 
which we find to be quite uncompetitive except at small values of $\Gamma/|K|$. 
The apparent convergence of KSL and U(1) energies with increasing field is 
due largely to spin polarization rather than to competition and is cut off 
by the phase transition to the trivial paramagnet (occurring, from Fig.~4(a) 
of the main text, at $B_c/g \mu_{\rm B} |K|$). Because the mean-field decoupling 
of the KSL \cite{KGammma_DMRG_YBKim_arX17} proceeds differently from 
Eq.~(\ref{Hmf}), this state cannot reduce to a U(1) state by the vanishing 
of off-diagonal expectation values. 

\begin{table}[b]
\centering
\begin{tabular}{c||c|c|c|c}
cluster size$\,$ & $E_{\rm ED}$ & $E_{\rm v}$ & $\,$relative error$\,$ & 
$\,$overlap \\ \hline
8 sites & $\, -0.6763 \,$ & $\, -0.6651 \,$ & 1.66\% & $\,$98.84\% \\
16 sites &  $-0.6476$  &  $-0.6181$  & 4.56\% & $\,$88.22\% \\
18 sites &  $-0.6533$  &  $-0.6223$  & 4.74\% & $\,$89.44\% \\
\end{tabular}
\caption{Ground-state energies obtained from ED and variational calculations 
on three different clusters, relative errors and wave-function overlaps. 
Energies are quoted in units of $|K|$.}\label{tab:eo}
\end{table}

\subsubsection{Benchmarking VMC by ED}

\begin{figure}[t]
\includegraphics[width=8.5cm]{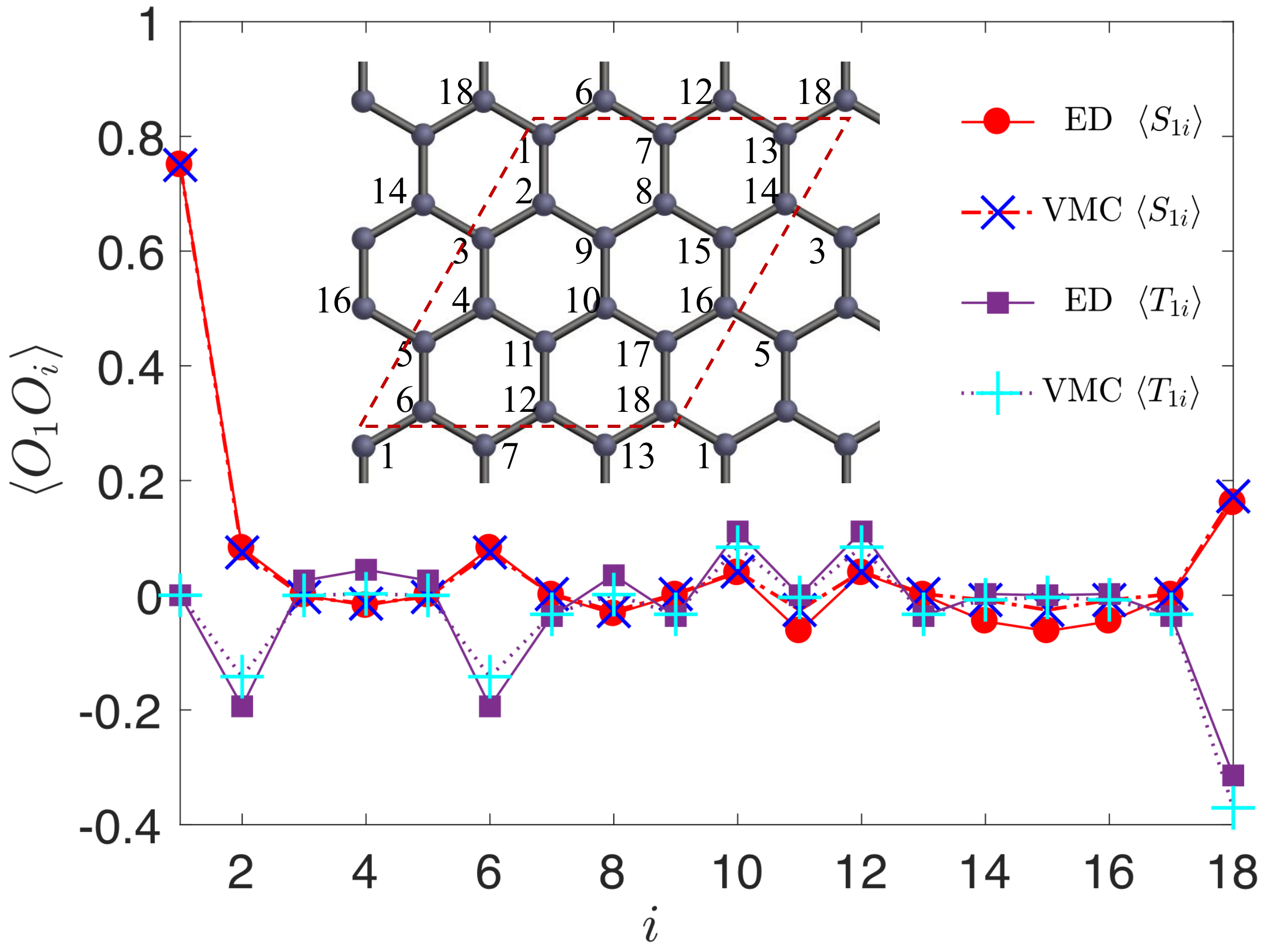}
\caption{Spin correlation functions, $S_{1i} = \langle \pmb S_1 \! \cdot \! 
\pmb S_i \rangle$, and off-diagonal symmetric correlation functions, $T_{1i} 
 = \langle S_1^x S_i^y + S_1^yS_i^x \rangle + \langle S_1^yS_i^z + S_1^zS_i^y 
\rangle + \langle S_1^zS_i^x + S_1^xS_i^z \rangle$, between site 1 and all 
other sites $i$ in the 18-site (3$\times$3 unit-cell) cluster.}  
\label{fig:corrs}    
\end{figure}

All of the VMC calculations we use to establish the magnetic order parameter 
and phase diagram for different field directions (Figs.~3 and 4 of the main 
text), and the corresponding spinon dispersions and gaps shown in Sec.~S3, 
are performed on systems of 8$\times$8 2-site unit cells (i.e.~128 sites).
We conclude this subsection by commenting on the benchmarking of these 
calculations, and of the optimal wave functions we construct, by comparison 
with exact-diagonalization calculations. We consider only our variational 
wave functions with no spinon pairing. We have performed ED on clusters 
of 2$\times$2, 2$\times$4, and 3$\times$3 unit cells, meaning systems with 
8, 16, and 18 sites. For these system sizes, variational calculations can 
be performed in full without resort to MC methods; because the efficacy of 
MC sampling methods is not in question, the comparison therefore serves to 
benchmark our variational procedure. In addition to the energies of these 
systems, we have computed the overlap, $\langle \psi_{\rm v}|\psi_{\rm ED} 
\rangle$, of the variational and ED wave functions to test their common 
content. We have also calculated the conventional and symmetric off-diagonal 
spin correlation functions, which are observables reflecting the spin state 
of both systems. As a result of the small ED system sizes, the zigzag 
magnetization is zero, and thus we cannot benchmark the magnetic order by 
this method (from VMC we find that system sizes of at least 4$\times$4 unit 
cells are required for finite $|\pmb M|$). The energies and overlaps shown 
in Table \ref{tab:eo} indicate a very close agreement, at the 90\% level 
for 16- and 18-site systems and the 99\% level on the 8-site system. The 
correlation functions, shown for the 18-site system in Fig.~\ref{fig:corrs}, 
demonstrate that the optimized variational states we have constructed do 
indeed capture all of the primary properties of the magnetic state of the 
system. 

\subsection{Parameters for modelling $\alpha$-RuCl$_3$}
\label{s1d}

It is necessary here to comment on the parameters required to model the 
physical properties of the $\alpha$-RuCl$_3$ system and on our choice of 
minimal model. The ($J, K, \Gamma, \dots$) parameter set appropriate for 
$\alpha$-RuCl$_3$ has been the subject of significant controversy, with 
not only the magnitudes but the signs and indeed the very presence of the 
different possible parameters being strongly contested. An excellent 
compilation, and the most comprehensive discussion to date, are provided 
in the recent study of Ref.~\cite{rjav}. 

These authors show 3 scenarios for zigzag magnetic order in the zero-field 
ground state of the classical $K$-$J$-$\Gamma$ model \cite{rjav}. Of these, 
only one is consistent with the highly anisotropic magnetization measured 
in $\alpha$-RuCl$_3$ \cite{Johnson_PRB_92}. This result, which has been 
interpreted as an effective $c$ axis $g$ factor of only 0.4, can in fact 
be taken as evidence for a large and negative $K$ combined with a larger 
and positive $\Gamma$. Noting that the $J$ values invoked are small, our 
current understanding of the situation is that the magnetization and the 
signs and magnitudes of $K$ and $\Gamma$ are important in $\alpha$-RuCl$_3$, 
whereas the zigzag order is the consequence of weak $J$ terms and thus is a 
relatively minor issue. Different authors have proposed several different 
combinations of possible $J$ terms to achieve zigzag order \cite{Winter_2016, 
Winter_2017}. The difficulty in deciding among these is compounded by the 
problem of extracting the properties of the quantum model from a knowledge of 
the classical one: as examples, we believe that the multi-$Q$ state obtained 
in Ref.~\cite{rjav} under the physically reasonable ($K$,$\Gamma$) scenario 
with a weak ferromagnetic (FM) nearest-neighbor $J$ term, and indeed any 
incommensurate-$Q$ state, would not be present in a quantum model. 

As stated in the main text, a linear spin-wave treatment based on this minimal 
$K$-$\Gamma$ model, with parameters $K < 0$, $\Gamma > 0$, and $\Gamma > |K|$, 
was found to provide a good fit to the spectrum of gapped (anisotropic) spin 
waves measured at zero field \cite{Ru_neutron_Wen_L17}. 
Our variational treatment 
of the zigzag magnetic order within this model was motivated by the result 
\cite{KJGamma_L14} 
that a zigzag-ordered ground state is favored 
within the single-$Q$ approximation. Subsequent analysis \cite{rjav} has shown 
that the classical model does not in fact support zigzag order without FM 
Heisenberg interactions. We stress again that rather little is known about 
the fully quantum $K$-$\Gamma$ model. 

The observation most important for the present work is the following. To 
include a FM Heisenberg interaction within the slave-fermion framework, it 
is most transparent to express this in the form 
\beq
\!\!\!\!\! - \pmb {S}_{i} \! \cdot \! \pmb {S}_{j} = - {\textstyle \frac{1}{4}} 
(C_i^\dag \pmb \sigma C_j \! \cdot \! C_j^\dag \pmb\sigma C_i + C_i^\dag \pmb 
\sigma \bar C_j \! \cdot \! \bar C_j^\dag \pmb\sigma C_i), \label{triplet_DC}
\eeq
which is identical to Eq.~(\ref{singlet_DC}) up to a different constant. 
Because the vector parameters $\langle C_i^\dag \pmb \sigma C_j \rangle$ 
and $\langle C_i^\dag \pmb \sigma \bar C_j \rangle$ contain only combinations 
of the $t^\gamma_{0,1,2,3}$ and $\Delta^\gamma_{0,1,2,3}$ terms which are already 
present in the Hamiltonian of Eq.~(\ref{Hmf}), the inclusion of a weak $J_1$ 
\cite{rjav} or $J_3$ term \cite{Winter_2016,Winter_2017} leads only to a 
small quantitative rescaling of the parameters of the system and cannot cause 
qualitative alterations to the nature of the spin-liquid state. We stress that 
this statement includes the issue of possible spinon-pairing terms discussed in 
Secs.~\ref{sec:u1orz2} and \ref{sec:VMC}: while these may indeed become more 
favorable for large changes to the input parameters, they cannot appear as a 
direct consequence of very small $J_1$ or $J_3$ terms. Thus all the primary 
conclusions of our variational analysis are robust against small changes 
to the model parameters and this is the sense in which we assert that the 
minimal $K$-$\Gamma$ model is fully representative of the extended Kitaev 
system in the general parameter space relevant to $\alpha$-RuCl$_3$. 

\section{Dirac cones: sign of mass and Chern number}

In this section we present a complete analysis of the combined symmetries of 
the honeycomb lattice and the spin-orbit-coupled spin sector, which act to 
protect the Dirac cones in the mean-field dispersion of Eq.~(2) of the main 
text. For simplicity, we consider the case with $t_1 = 1$, $t_0 = t_2 = 0$, 
and $|\pmb B| = 0$, where the mean-field Hamiltonian becomes 
\beq\label{MF0}
H_{\rm mf} = -i \sum_{\langle i,j\rangle \in(\alpha\beta)\gamma} C_i^\dagger (\sigma_\alpha
 + \sigma_\beta) C_j.
\eeq
We expand Eq.~(\ref{MF0}) in Fourier space and consider the Dirac cones 
centered at the different points shown in Fig.~2(b) of the main text. 

\subsection{Points $K$ and $K'$}

The wave vector $\pmb K$ is invariant under the action of a $C_{3v}$ group. 
Because $C_{3v}$ is a symmetry of the mean-field Hamiltonian (\ref{MF0}), 
the Fourier component $H_{\pmb K}$ must be invariant under the action of 
$C_{3v}$. Due to the spin-orbit coupling, the full symmetry is a combination 
of spin and sublattice operations, for which an explicit expression can be 
obtained by considering the Hamiltonian component 
\beq
H_{\pmb K} & = & - i C_{\pmb K,A}^\dag \left[ (\sigma_x + \sigma_y) \omega + 
(\sigma_y + \sigma_z) \right.\nonumber \\ & & \ \ \ \ \ \ \ \ \ \ \ \ \ \ 
\ \ \ \ \left. + (\sigma_x + \sigma_z) \omega^2 \right] C_{\pmb K,B} + {\rm h.c.} 
\nonumber \\ & = & 
C_{\pmb K}^\dag \left[ - \mu_y \otimes \sigma_x + \left( {\textstyle 
\frac{\sqrt3}{2}} \mu_x + {\textstyle \frac12} \mu_y \right) \otimes 
\sigma_y \right. \nonumber\\ & & \ \ \ \ \ \ \ \ \ \ \ \left. + \left(- 
{\textstyle \frac{\sqrt3}{2}} \mu_x + {\textstyle \frac12} \mu_y \right) 
\otimes \sigma_z 
\right] C_{\pmb K},\label{HK}
\eeq
in which $\omega = e^{i{2\pi\over3}}$, $C_{\pmb K,\alpha}^\dag = (c_{\pmb K,\alpha,\up}^\dag 
\;\, c_{\pmb K,\alpha,\dn}^\dag)$, where the index $\alpha = A,B$ represents the two 
sublattices of the honeycomb system, and $C_{\pmb K}^\dag = (C_{\pmb K,A}^\dag \;\, 
C_{\pmb K,B}^\dag)$. The Pauli-matrix operators $\mu_m$, with $m = x, y, z$, 
act on the sublattice degrees of freedom while the operators $\sigma_m$ 
act on the spin degrees of freedom. 

\begin{figure}[t]
\includegraphics[width=8cm]{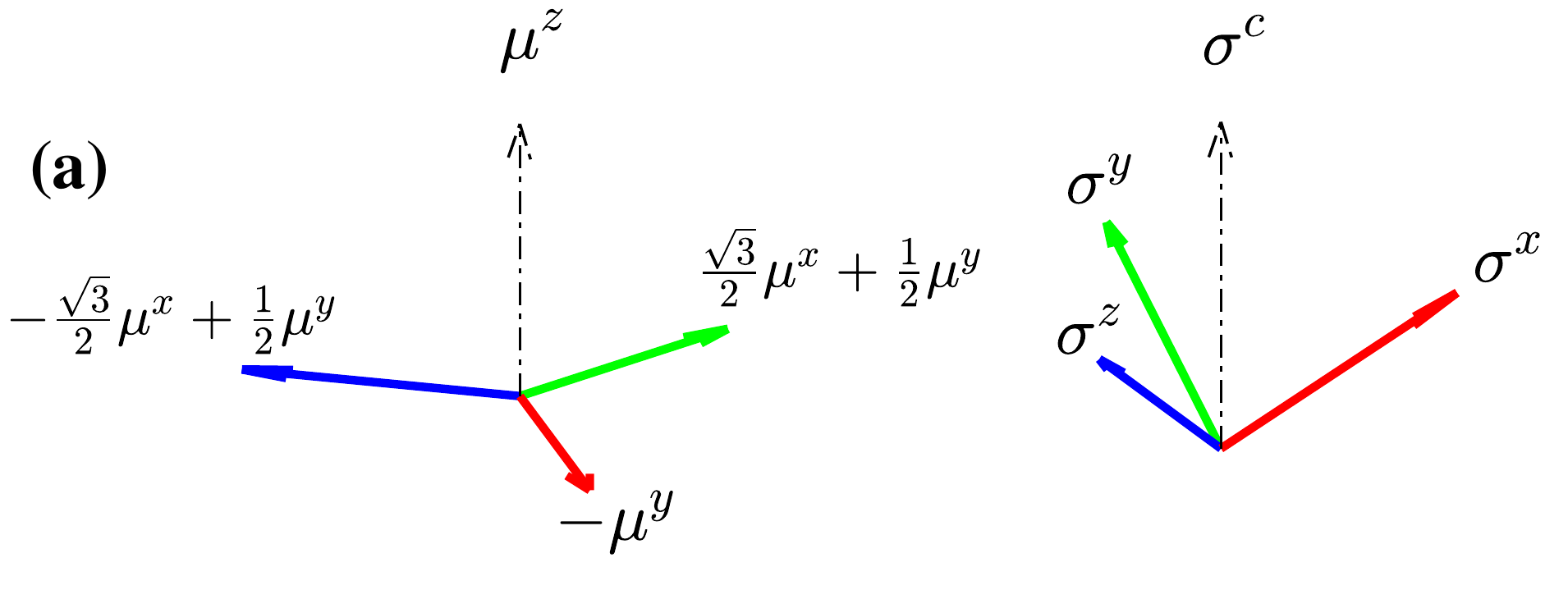}
\includegraphics[width=8cm]{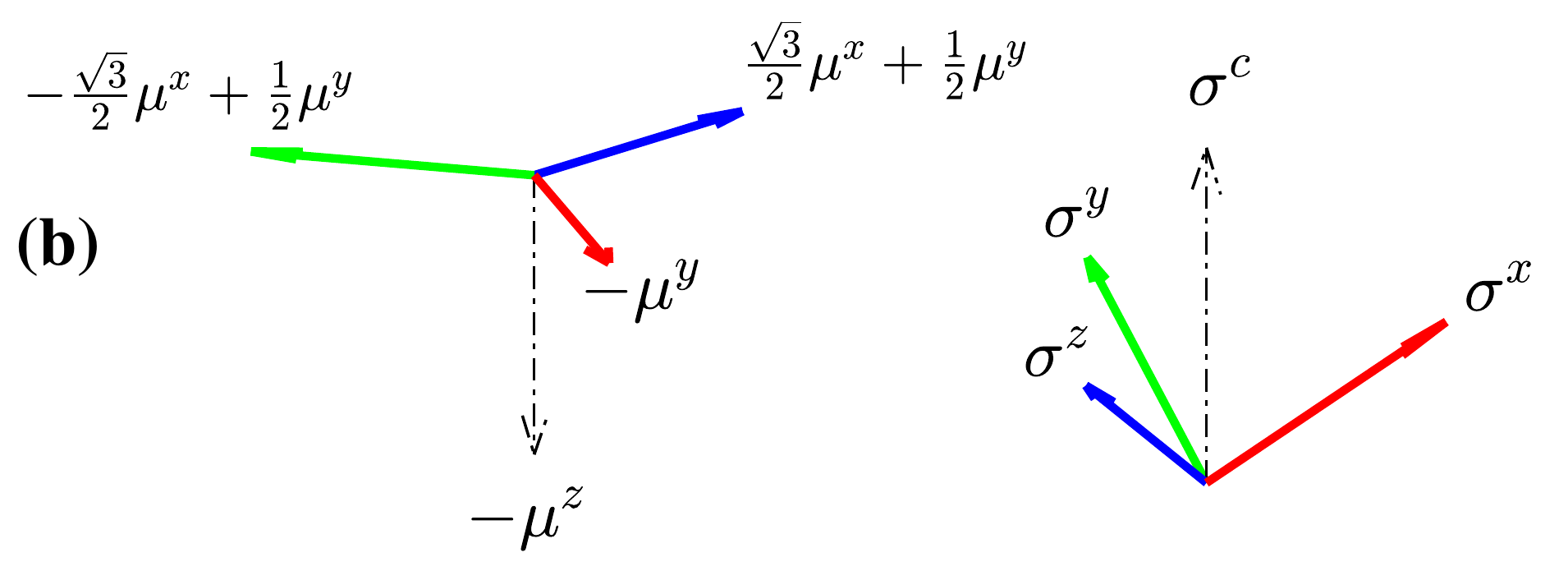}
\caption{Schematic representation of the operator content of the combined 
sublattice and spin symmetries contained in the minimal model of Eq.~(\ref{HK})
for the $K$ point of the Kitaev honeycomb system in the geometry presented by 
the materials Na$_2$IrO$_3$ and $\alpha$-RuCl$_3$. The left panels show the 
three directions in the sublattice space connected by the $\mu_x$ and $\mu_y$ 
operations, to which $\mu_z$ is orthogonal, while the right panels show the 
three directions in spin space governing the action of $\sigma_x$, $\sigma_y$, 
and $\sigma_z$. Vectors of the same color appear in the same terms in the 
Hamiltonians, $H_{\pmb K}$ in panel (a) and $H_{\pmb K'}$ in panel (b). The 
dot-dashed lines marks the axis of the $C_3$ symmetry, which causes a 
cyclic permutation of both sets of vectors.}
\label{fig:Vects}    
\end{figure}

A graphical understanding of the operator content of Eq.~(\ref{HK}) is provided 
in Fig.~\ref{fig:Vects}(a). Each of the three terms in the square brackets is 
a direct product of two 2$\times$2 matrices, $\mu_n \otimes \sigma_l$,
which we express in the form $\mu_{\pmb n} \otimes \sigma_{\pmb l}$, with 
$\mu_{\pmb n} = \pmb \mu \! \cdot \! {\hat{\pmb n}}$ and $\sigma_{\pmb l} = \pmb 
\sigma \! \cdot \! \hat{\pmb l}$. $\hat{\pmb n}$ and $\hat{\pmb l}$ are each 
a set of unit vectors determining the action of the operators, with 
$\hat{\pmb l}$ being simply the spin basis vectors ($\hat x, \hat y, 
\hat z$) encountered in Fig.~1(a) of the main text. As shown in 
Fig.~\ref{fig:Vects}(a), the corresponding $\hat {\pmb n}$ vectors form 
a non-orthogonal set spanning a plane; although the $\mu_m$ matrices operate 
in a somewhat abstract sublattice space, the hopping matrix elements related 
by $\mu_x$ and $\mu_y$ are subject to the symmetries of the honeycomb plane, 
and the $\mu_z$ operator is orthogonal to both. Vectors of the same color in 
Fig.~\ref{fig:Vects}(a) are connected by the same term in Eq.~(\ref{HK}). The 
$C_3$ rotation operation is a simultaneous cyclic permutation of the three 
$\hat{\pmb n}$ vectors and the three $\hat{\pmb l}$ vectors, while a mirror 
operation is a simultaneous exchange of two out of each set of basis vectors. 
Thus it is clear that the matrix operators representing the generators of 
the $C_{3v}$ symmetry may be written as 
\beq\label{C3vK}
C_3 & = & e^{-i{\mu_z\over2}{2\pi\over3}} \otimes e^{-i{\sigma_c\over2}{2\pi\over3}}, 
\label{C3K} \\
M_z & = & \left( {\textstyle \frac12} \mu_x + {\textstyle \frac{\sqrt3}{2}} 
\mu_y \right) \otimes {\textstyle \frac{1}{\sqrt2}} (\sigma_x - \sigma_y),
\eeq
with $\sigma_c = {1\over\sqrt3}(\sigma_x + \sigma_y + \sigma_z)$, which is 
contained in Eq.~(3) of the main text. The $C_{3v}$ symmetry is non-Abelian 
and protects the twofold energy-level degeneracy at the $K$ point, 
which gives rise to a Dirac cone. 

Because $\pmb K' = - \pmb K$, the Hamiltonian at $K'$ can be obtained by 
exchanging $\omega$ with $\omega^2$, whence 
\beq
H_{\pmb K'} & = & C_{\pmb K'}^\dag \left[ - \mu_y \otimes \sigma_x + \left(- 
{\textstyle \frac{\sqrt3}{2}} \mu_x + {\textstyle \frac12} \mu_y \right) 
\otimes \sigma_y \right. \nonumber\\& & \ \ \ \ \ \ \ \ \ \ \ \left. + \left( 
{\textstyle \frac{\sqrt3}{2}} \mu_x + {\textstyle \frac12} \mu_y \right) 
\otimes \sigma_z \right] C_{\pmb K'},\label{HKp}
\eeq
and the generators of $C_{3v}$ at $K'$ are represented as 
\Beq\label{C3vK'}
C_3' & = & e^{ i{\mu_z\over2}{2\pi\over3}} \otimes e^{-i{\sigma_c\over2}{2\pi\over3}}, 
\label{C3K'} \\
M_z' & = & \left(- {\textstyle \frac12} \mu_x + {\textstyle \frac{\sqrt3}{2}} 
\mu_y \right) \otimes {\textstyle \frac{1}{\sqrt2}} (\sigma_x - \sigma_y),
\Eeq
which is represented in Fig.~\ref{fig:Vects}(b)  and constitutes the other 
half of Eq.~(3) of the main text. 

Next we note that the mean-field Hamiltonian (\ref{MF0}) has spatial-inversion 
symmetry, $C_i= \{E,P\}$, which is a subgroup of the full symmetry group, 
$D_{3d}$. The inversion operation reverses the sign of the wave vector, 
i.e.~$\hat P \pmb k = -\pmb k$. When acting on the matrix elements of the 
Hamiltonian, this is equivalent to reversing the bond direction, which is 
the same as permuting the sublattice indices. This result can be observed by 
considering the relation between $H_{\pmb K'}$ and $H_{\pmb K}$, which yields
$$ \hat P H_{\pmb K} {\hat P}^{-1} = H_{\pmb K'} = C_{\pmb K'}^\dag \mathcal H_{\pmb K'} 
C_{\pmb K'} = C_{\pmb K'}^\dag \mu_y \mathcal H_{\pmb K} \mu_y C_{\pmb K'}, $$
where $\mathcal H_{\pmb K}$ denotes the matrix Hamiltonian operator within 
$H_{\pmb K}$ and $\mu_y \equiv \mu_y \otimes I$ denotes the combined operator 
with sublattice and spin components. Thus for any general momentum, $\pmb k$, 
the relation $\hat P H_{\pmb k} {\hat P}^{-1} = H_{- \pmb k}$ means that $H_{\pmb k}$ 
has the property 
\beq\label{Inv}
\mathcal H_{-\pmb k} = \mu_y \mathcal H_{ \pmb k}~\mu_y.
\eeq
Further, because the Hamiltonian contains no intra-sublattice (second-neighbor) 
spinon hopping terms, and thus contains only $\mu_x$ and $\mu_y$, spatial 
inversion can be used to deduce the additional property 
\beq\label{muz}
\mu_z \mathcal H_{\pmb k} ~\mu_z = -\mathcal H_{\pmb k}.
\eeq

From these symmetries of the Hamiltonian, it is possible to read out two 
important pieces of information, namely (i) which types of perturbation will 
break the total symmetry, causing a gap to open in the Dirac cones, and (ii) 
if the Dirac cones are gapped, what the resulting Chern number should be. 

To obtain this information, we focus on wave vectors near the two Dirac points. 
When a Dirac cone is gapped, a half-quantized Chern number ${\mathcal C}
 = \pm {1\over2}$ is obtained \cite{Jackiw_1984,Haldane_1988,WHA_2010}. The 
sign of the Chern number is also said to be the sign of the mass. To analyze 
the total Chern number of the Dirac cones, we define the matrix Hamiltonian 
operators 
\Beq
\delta \mathcal H(\delta \pmb k) & = & \mathcal H_{\pmb K+\delta \pmb k} - 
\mathcal H_{\pmb K},\\
\delta \mathcal H'(\delta \pmb k) & = & \mathcal H_{\pmb K'+\delta \pmb k} - 
\mathcal H_{\pmb K'},
\Eeq
and keep only those terms linear in $\delta \pmb k$ at small $|\delta \pmb k|$. 
This approximation is in general reliable because the primary contributions to 
the Chern number in the presence of a mass term are from states very close to 
the Dirac point. By substituting Eq.~(\ref{muz}) into (\ref{Inv}), we deduce 
that $\mathcal H_{-\pmb k} = - \mu_x \mathcal H_{ \pmb k} \mu_x$
and hence 
\beq
\mu_x \delta \mathcal H(\delta \pmb k) \mu_x & = & - (\mathcal H_{- \pmb K
 - \delta\pmb k} - \mathcal H_{- \pmb K}) \nonumber 
\\ & = & - (\mathcal H_{\pmb K' - \delta\pmb k} - \mathcal H_{\pmb K'}) \nonumber 
\\ & = & - \delta \mathcal  H'(-\delta \pmb k) \; = \; \delta \mathcal H' 
(\delta \pmb k), 
\label{KK'}
\eeq
where the last equality follows from the assumption of linearity. 

As noted in the main text, there are two types of mass term which break 
the $C_{3v}$ symmetry and gap the pair of Dirac cones at $K$ and $K'$, 
namely a sublattice chemical potential, $\lambda \mu_z$, and a Zeeman field 
term, $g\mu_B\sigma_c$. Without knowing any details of $\mathcal H_{\pmb K + 
\delta \pmb k}$, it is clear from their differing operator structures that 
the two mass terms yield different total Chern numbers. 

To demonstrate that the sublattice potential term gives a trivial total 
Chern number, we show that the $\lambda \mu_z$ term has different signs 
for the mass term at the two Dirac cones. At the $K$ point, the perturbed 
Hamiltonian giving the dispersion a massive Dirac-cone form is 
$$h(\delta\pmb k, \lambda) = \delta \mathcal H(\delta\pmb k) + \lambda\mu_z.$$ 
From Eq.~(\ref{KK'}), at the $K'$ point one has 
\Beq
h'(\delta\pmb k, \lambda) & = & \delta \mathcal H'(\delta\pmb k) + \lambda 
\mu_z \\ & = & \mu_x \delta [\mathcal H(\delta {\pmb k}) - \lambda \mu_z] \mu_x 
\\ & = & \mu_x h (\delta \pmb k, -\lambda) \mu_x.
\Eeq
Because a global unitary transformation such as $\mu_x$ does not change the 
topological properties of the ground state, the Chern number of $h' (\delta 
\pmb k, \lambda)$ must equal that of $h (\delta \pmb k, -\lambda)$, which 
cancels the contribution from $h (\delta \pmb k, \lambda)$, and therefore
the total Chern number contributed by the mass term $\lambda \mu_z$ is zero. 

To demonstrate that the Zeeman term gives a nontrivial total Chern number, we 
compare the two gapped Dirac Hamiltonians
\Beq
h (\delta \pmb k, B_c) & = & \delta \mathcal H (\delta \pmb k) + g \mu_B B_c 
\sigma_c, \\ h' (\delta \pmb k, B_c) & = & \delta \mathcal H' (\delta \pmb k)
 + g \mu_B B_c \sigma_c.
\Eeq
Because $\mu_x$ commutes with $\sigma_c$, from (\ref{KK'}) we have 
\[ h'(\delta\pmb k, B_c) = \mu_x (\delta \mathcal H(\delta\pmb k) + g\mu_BB_c
\sigma_c)\mu_x = \mu_x h(\delta\pmb k, B_c) \mu_x.
\]
From the fact that $h(\delta \pmb k, B_c)$ and $h'(\delta \pmb k, B_c)$ are 
related by a global unitary transformation, they must always have the same 
Chern number, i.e.~$+1/2$ or $- 1/2$. As a consequence, the total Chern number 
of the two Dirac cones contributed by the mass term $g \mu_B B_c \sigma_c$ is 
either 1 or $-1$. 

\begin{figure*}[t]
\includegraphics[width=8.0cm]{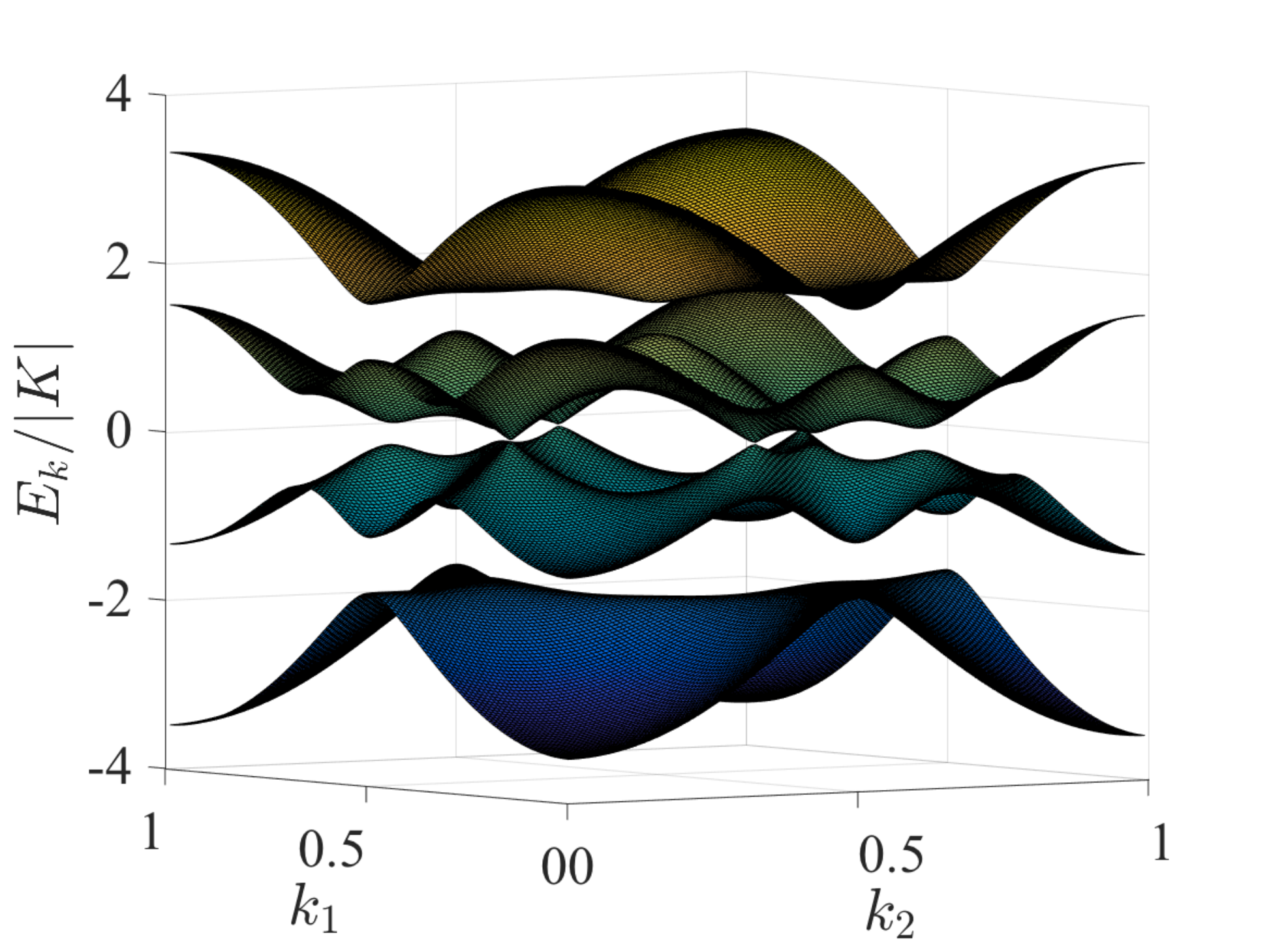}\hskip5mm 
\includegraphics[width=8.0cm]{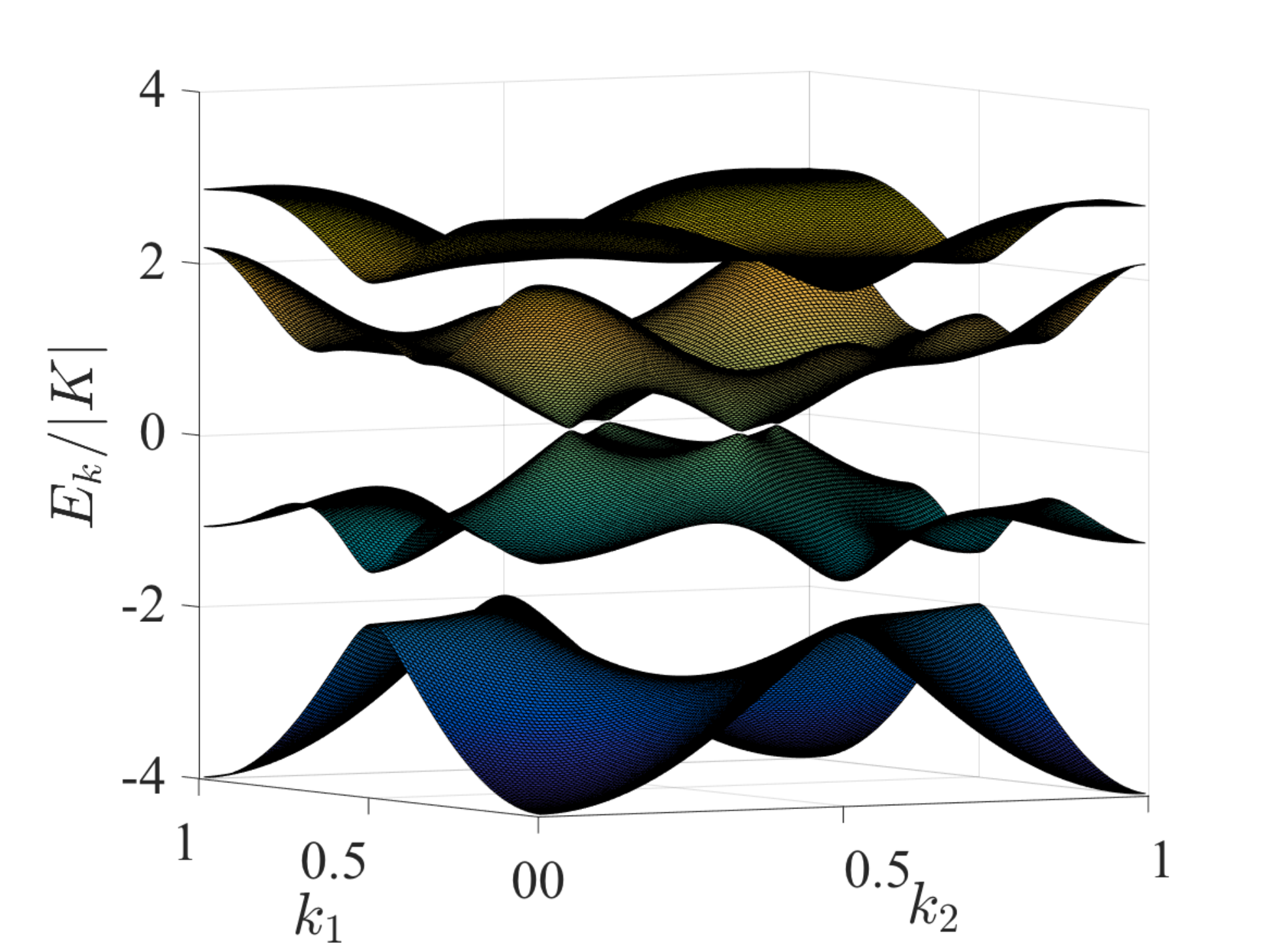} 
\centerline{\qquad\qquad (a) \qquad\qquad\qquad\qquad\qquad\qquad\qquad
\qquad\qquad\qquad\qquad\qquad\qquad\qquad (b)} 
\includegraphics[width=8.0cm]{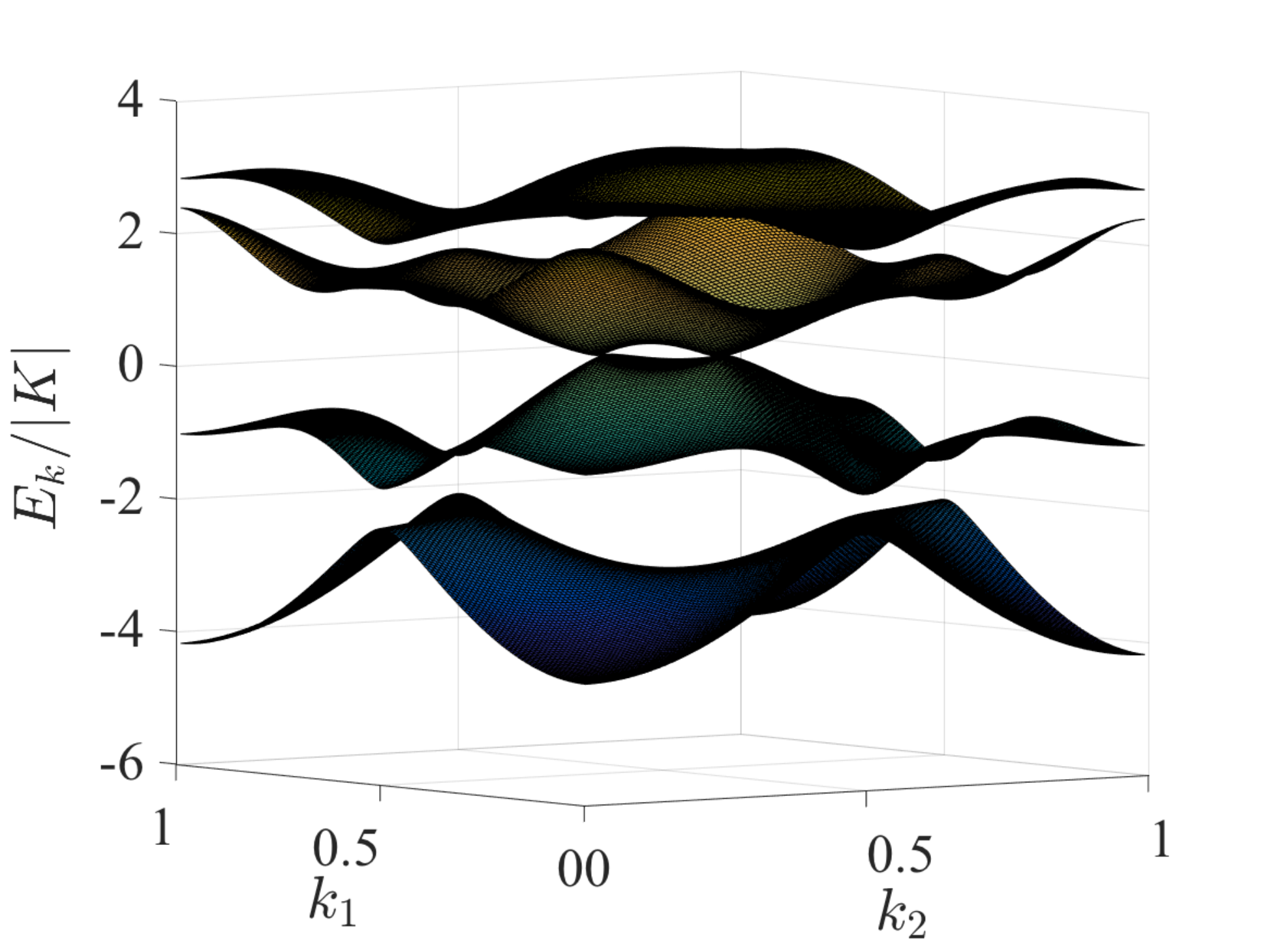}\hskip5mm  
\includegraphics[width=8.0cm]{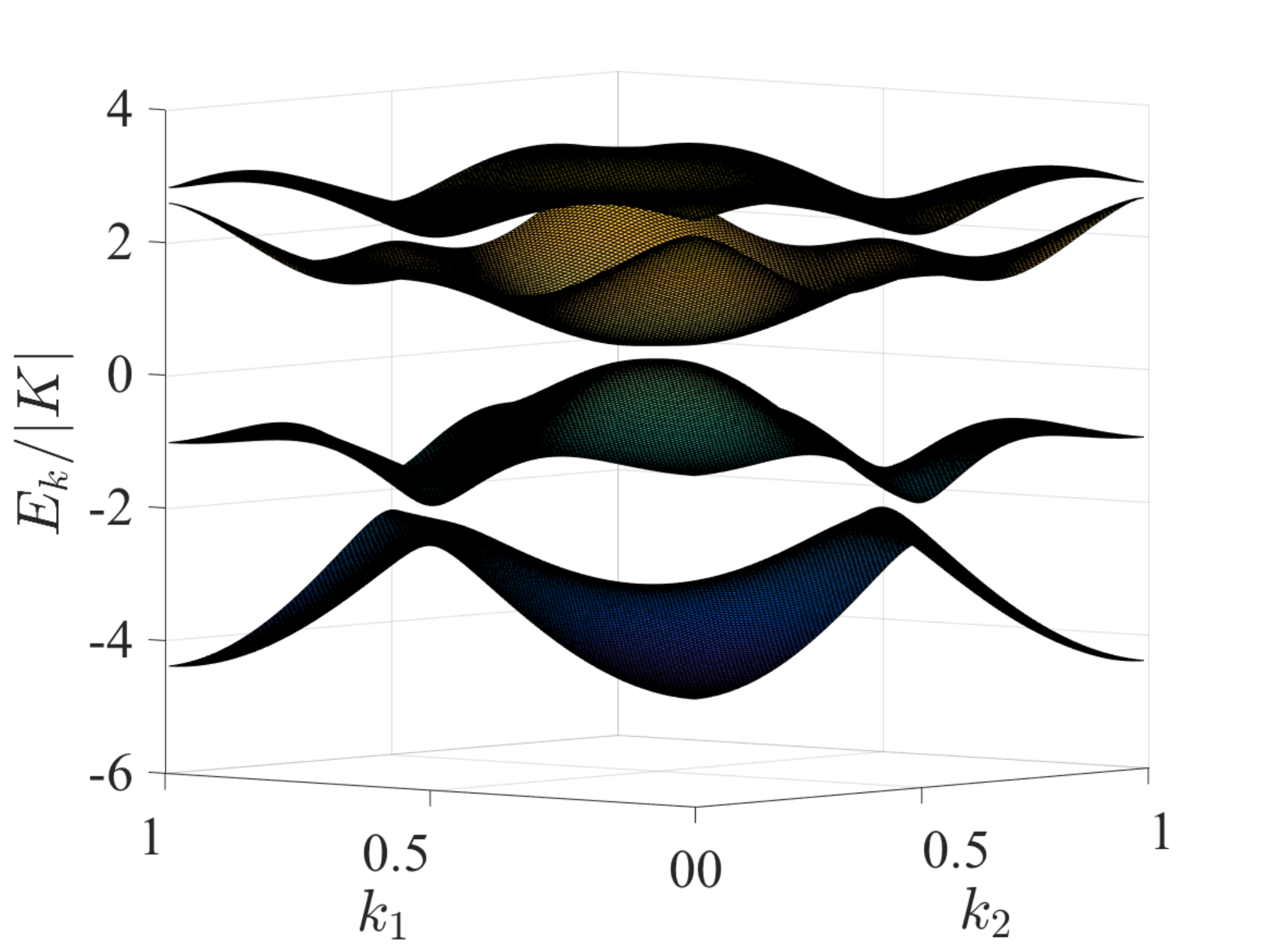} 
\centerline{\qquad\qquad (c) \qquad\qquad\qquad\qquad\qquad\qquad\qquad
\qquad\qquad\qquad\qquad\qquad\qquad\qquad (d)} 
\caption{Spinon dispersion relations of the Dirac spin liquid induced by 
a magnetic field $\pmb B \parallel \hat z$. (a) When $g\mu_BB_z/|K| = 0.16$, 
there are two pairs of Dirac cones located close to the points $K_x,K_y$ and 
$K_x',K_y'$ in Fig.~2(b) of the main text. (b) When the field is increased, 
the nodes from $K_x$ and $K_y$, and those from $K_x'$ and $K_y'$, move toward 
each other, as illustrated here for $g\mu_BB_z/|K| = 1$. (c) When the field 
reaches a critical value, $g\mu_BB_z^c/|K| = 1.25$ in this case, the cones 
from each pair merge into a single gapless point with semi-Dirac dispersion. 
(d) When the field is increased further, as shown here for $g\mu_BB_z/|K| = 
1.5$, a gap opens and the system enters the trivial gapped phase.}
\label{fig:Bz}    
\end{figure*}

\subsection{Points $K_x$ and $K_x'$}

Following the notation and logic of the previous subsection, the $K_x$ 
component of the Hamiltonian is 
\[
H_{\pmb K_x} = C_{\pmb K_x}^\dag \left[ \mu_x \otimes (- \sigma_y + \sigma_z) + \mu_y
\otimes (\sigma_y + \sigma_z) \right] C_{\pmb K_x}.
\]
Although the momentum $\pmb K_x$ has a relatively low symmetry, as is evident 
from Fig.~2(b) of the main text, the Hamiltonian at this point has an emergent 
non-Abelian $C_{4v}$ symmetry, whose two generators are 
\beq
C_4 & = & e^{-i {\mu_z\over2}{\pi\over2}} \otimes e^{i{\sigma_x\over2}{\pi\over2}}, 
\label{C4Kx} \\ M_x & = & \mu_x \otimes {\textstyle \frac{1}{\sqrt2}} 
(\sigma_y - \sigma_z), 
\eeq
as stated in Eq.~(4) of the main text. Similar to the treatment of the $K$ 
and $K'$ points, the Hamiltonian term $H_{\pmb K_x'}$ is related to $H_{\pmb K_x}$ 
by $\mathcal H_{\pmb K_x'} = \mu_y \mathcal H_{\pmb K_x} \mu_y$, whence 
\Beq
H_{\pmb K_x'} = C_{\pmb K_x'}^\dag \left[ -\mu_x \otimes (-\sigma_y + \sigma_z) + \mu_y
\otimes (\sigma_y + \sigma_z) \right] C_{\pmb K_x'},
\Eeq
and the generators of the emergent $C_{4v}$ symmetry are 
\Beq
C_4' & = & \mu_y C_4 \mu_y = e^{i {\mu_z\over2}{\pi\over2}} \otimes e^{i{\sigma_x\over2}
{\pi\over2}},\\
M_x' & = & \mu_y M_x \mu_y = - \mu_x \otimes {\textstyle \frac{1}{\sqrt2}} 
(\sigma_y - \sigma_z).
\Eeq
Following the logic applied at the $K$ and $K'$ points, a mass term $\lambda 
\mu_z$ gaps the Dirac cones with topologically trivial consequences while a 
term $g \mu_B B_x \sigma_x$ contributes a nonzero total Chern number of $1$ 
or $-1$. 

\begin{figure*}[t]
\leftline{
\includegraphics[width=6cm]{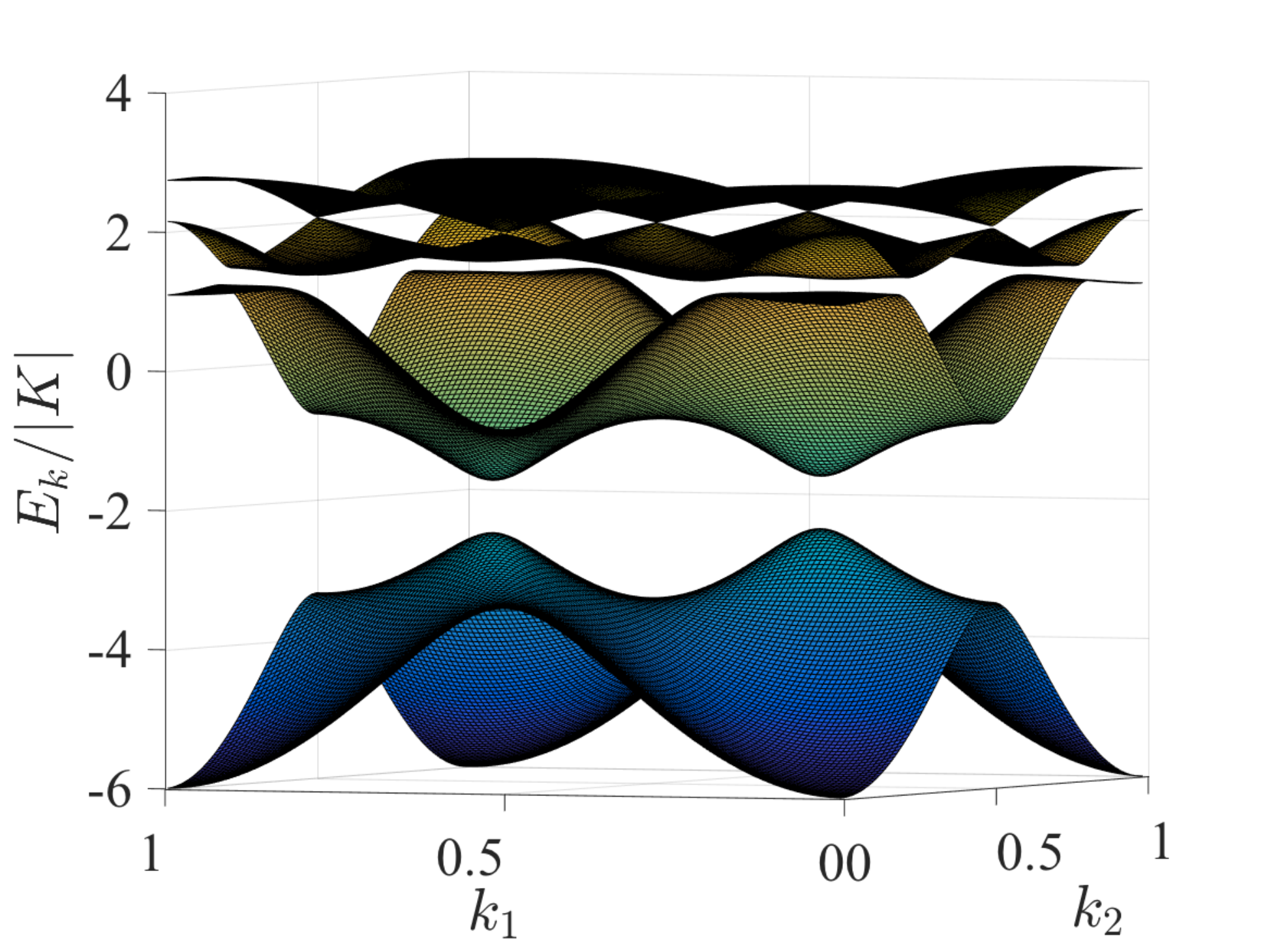} \hskip-4.2mm
\includegraphics[width=6cm]{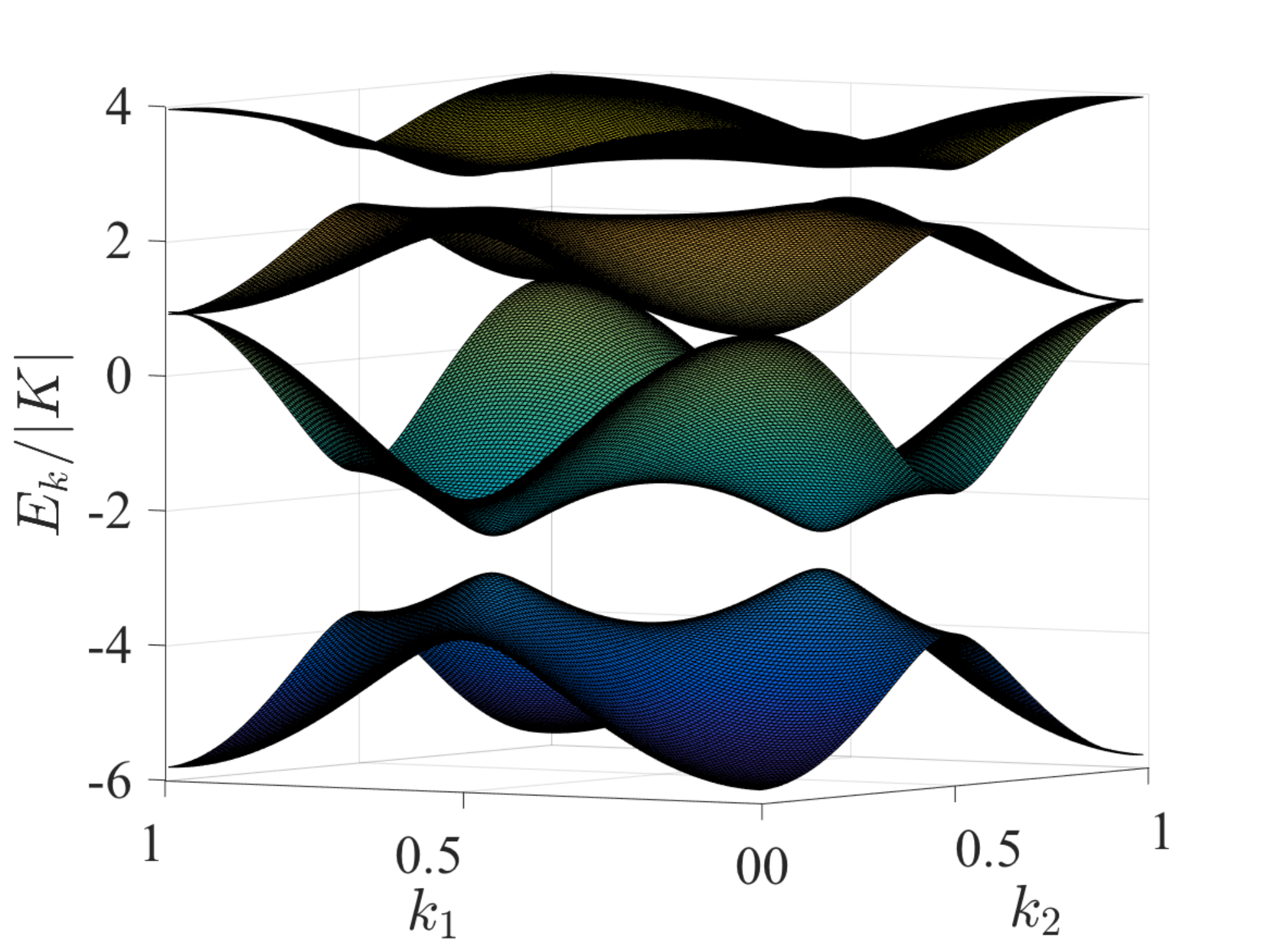} \hskip-4.2mm
\includegraphics[width=6cm]{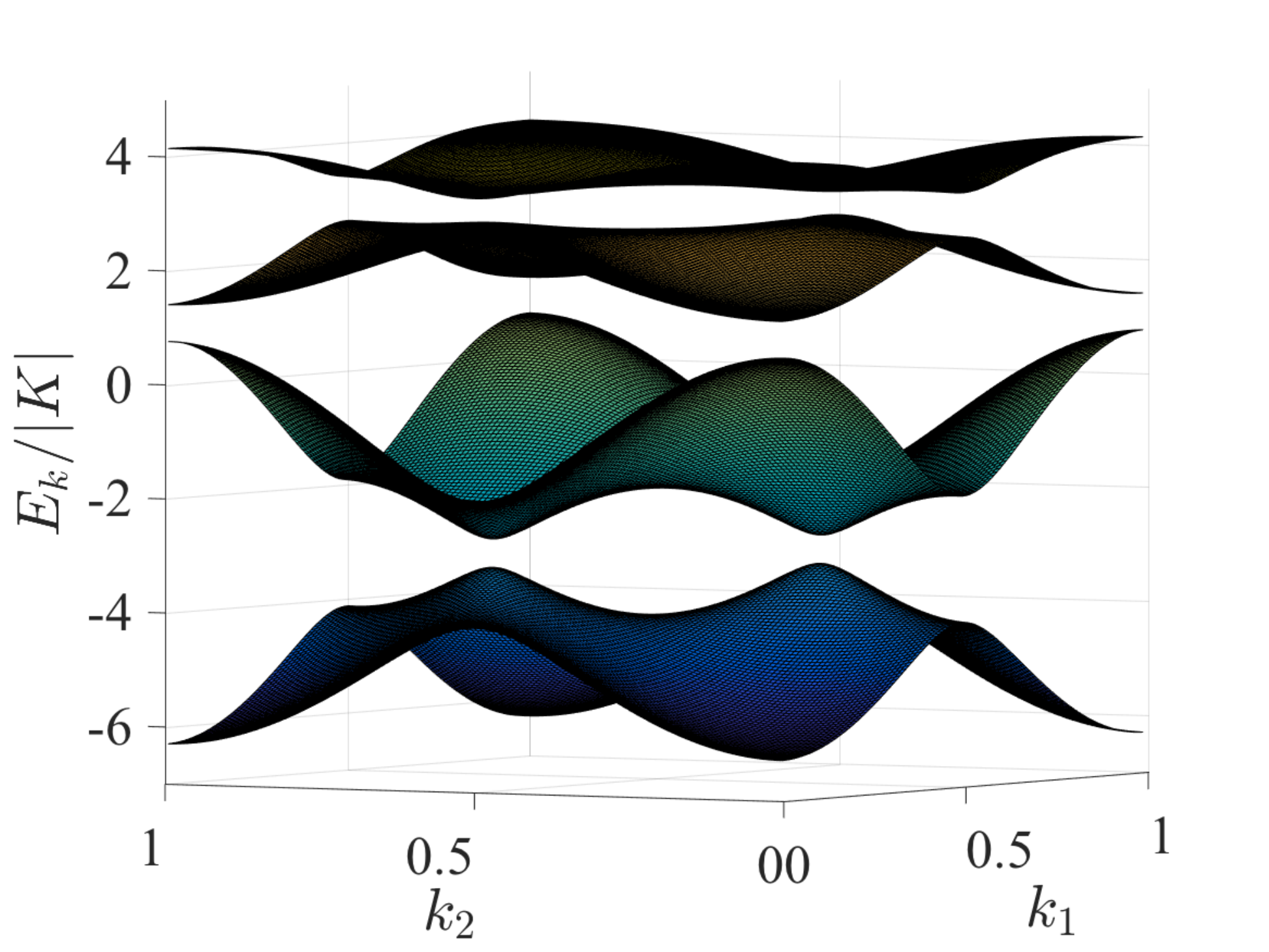}}
\centerline{(a) \qquad\qquad\qquad\qquad\qquad\qquad\qquad\qquad (b) 
\qquad\qquad\qquad\qquad\qquad\qquad\qquad\qquad (c)}
\caption{Spinon dispersion relations of the chiral spin liquid induced by a 
magnetic field $\pmb B$ applied in the direction $(\hat x + \hat y + \hat z)$.
(a) When $g\mu_B \pmb B/|K| = (1.4,1.4,1.4)$, the spinon band structure is 
fully gapped and when half-filled has total Chern number ${\mathcal C} = 2$. 
(b) When the field is increased to a critical value, $g\mu_B\pmb B^c/|K| = 
(1.8,1.8,1.8)$, the gaps close as semi-Dirac ``cones'' form. (c) When the 
field is increased further, as shown here for $g\mu_BB/|K| = (2.2,2.2,2.2)$, 
the gap reopens and the system enters the trivial gapped phase.}
\label{fig:Bxyz}    
\end{figure*}

However, the relative signs of the Chern numbers should be treated carefully. 
From Eq.~(\ref{C3K}), $\mu_z$ and $\sigma_c$ define the direction of the $C_3$ 
rotation. If either term is added to the Hamiltonian at $K$, the $C_{3v}$ 
symmetry is broken in the same way and thus both terms give the same sign of 
the mass for the resulting Dirac cone. By contrast, Eq.~(\ref{C4Kx}) shows 
that $\mu_z$ and $-\sigma_x$ give the same sign of the mass for the Dirac 
cones they gap at $K_x$. Alternatively stated, the Chern number generated by 
the mass $\sigma_c$ at the cones $K$ and $K'$ is the same as the Chern number 
generated by the mass $-\sigma_x$ at the cones $K_x$ and $K_x'$. Specifically, 
if $g \mu_B B_0 \sigma_c$ term contributes a total Chern number of $-1$ for 
the pair of Dirac cones at $K$ and $K'$, then $g \mu_B B_0 \sigma_x$ 
contributes a net Chern number of 1 for the pair of Dirac cones at $K_x$ 
and $K_x'$. The discussion for the other two pairs of cones, $K_y, K_y'$ 
and $K_z, K_z'$ is the same as the case $K_x, K_x'$ and will not be repeated 
here. These results underpin the content of Table I in the main text and 
the expression 
$$\mathcal C = {\rm sgn} (\pmb B \! \cdot \! \hat x) + {\rm sgn}(\pmb B \! 
\cdot \! \hat y) + {\rm sgn}(\pmb B \! \cdot \! \hat z) - {\rm sgn}(\pmb B 
\! \cdot \! \hat c)$$
deduced there. 

The symmetry analysis we have applied at the Dirac points is strictly valid 
only for the Hamiltonian at $\pmb B = 0$, whereas a finite magnetic field 
is expected to violate some of its symmetries. However, the conclusions 
drawn from these symmetry arguments remain valid if the relative field 
intensity, $g \mu_B |\pmb B|/|K|$, is small. This situation is also analogous 
to the case of graphene, where the $C_{3v}$ symmetry protects the Dirac cones 
at $K$ and $K'$. If a small strain acts to deform the graphene sheet, the 
$C_{3v}$ symmetry is no longer satisfied rigorously, but the Dirac cones survive 
with small shifts of their positions in momentum space \cite{GrapheneSt}. In 
the present analysis, the application of a magnetic field oriented in one of 
the three directions ($\hat x - \hat y$), ($\hat y - \hat z$), or ($\hat z
 - \hat x$) results in the Dirac cones at $K$ and $K'$ surviving, but with 
their positions shifted. In Sec.~S3 we demonstrate numerically the 
robustness of the Dirac cones over a finite range of field intensity, 
which reflects the fact that they are indeed symmetry-protected.

\section{Spinon dispersion relations}

Here we illustrate the form of the spinon dispersion relations obtained when 
the magnetic field is applied in different directions relative to the crystal 
axes. We assume that the zigzag magnetic order is suppressed by the action 
of the field. In general, the spinon band structure obtained at the mean-field 
level remains qualitatively unaltered by the Gutzwiller projection, although 
the band width and band gap are renormalized. The spinon dispersions shown 
in the figures to follow are computed from the mean-field Hamiltonian with 
variational parameters adopted from VMC calculations in which the energy of 
the trial ground state was optimized. 

\noindent
{\bf Field-induced Dirac spin liquid.} If $\pmb B$ is parallel to one of the 
directions $\hat x$, $\hat y$, $\hat z$, $(\hat x - \hat y)$, $(\hat y - \hat 
z)$, or $(\hat z - \hat x)$, there exists a field-induced gapless spin-liquid 
phase. In Fig.~\ref{fig:Bz} we show the spinon spectrum for the case $\pmb B 
\parallel \hat z$. It is clear that when the field intensity exceeds a lower 
critical value, required to suppress the zigzag ordered phase (Figs.~3 and 
4 of the main text), it induces a spin liquid with two pairs of Dirac cones 
[Fig.~\ref{fig:Bz}(a)]. Increasing the field causes the two cones in the 
left half of the Brillouin zone to move towards each other, while the pair 
in the right half behaves symmetrically [Fig.~\ref{fig:Bz}(b)]. This process 
continues until the cones of each pair merge to form a single gapless point, 
about which the dispersion is of semi-Dirac type [Fig.~\ref{fig:Bz}(c)]. A 
further increase in field beyond this critical value causes the two new 
semi-Dirac ``cones'' to become gapped [Fig.~\ref{fig:Bz}(d)].

\begin{figure}[b]
\includegraphics[width=8.5cm]{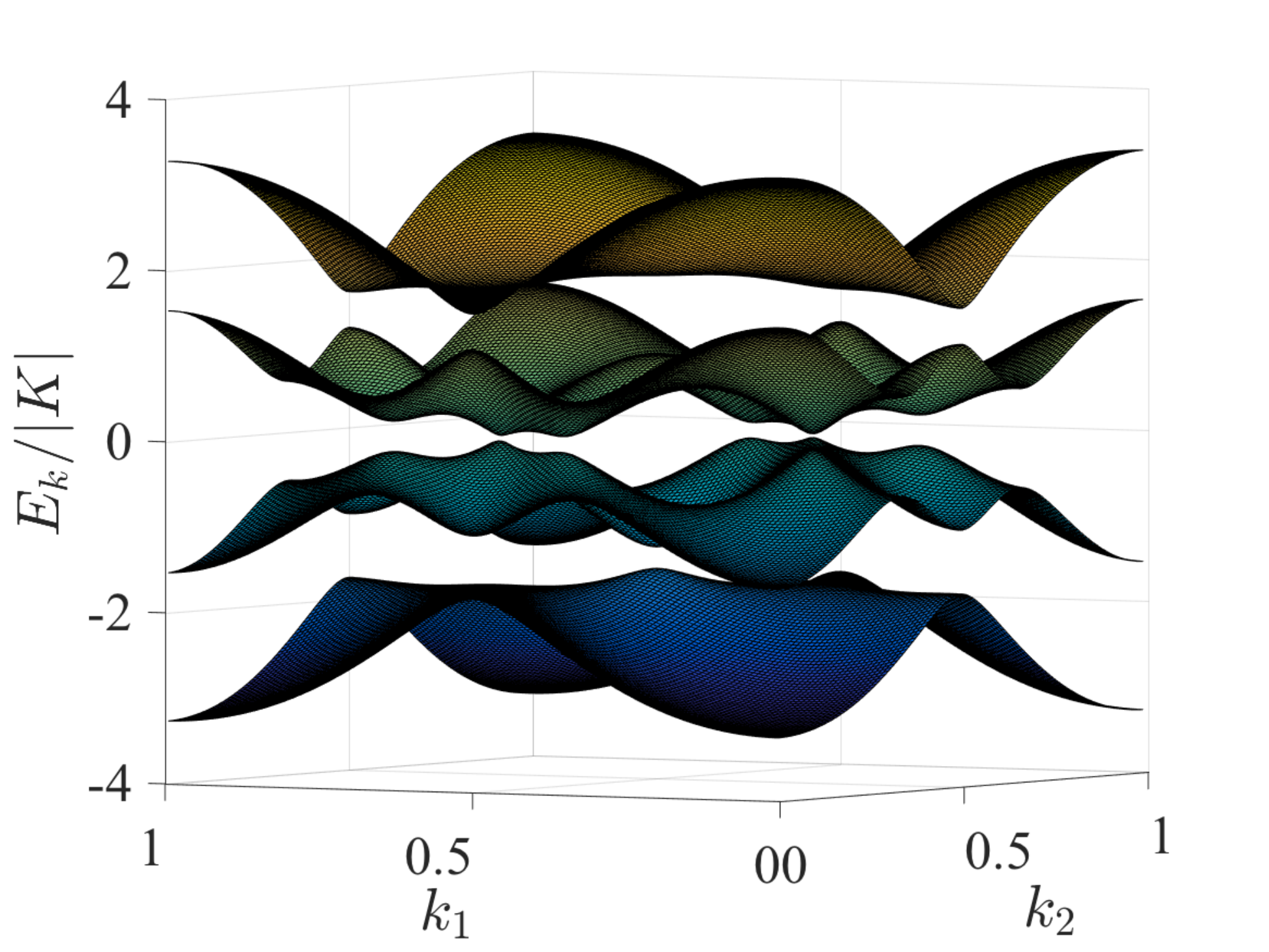}
\caption{Spinon dispersion for $g \mu_B \pmb B /|K| = (0.1, 0.1, -0.2)$. 
Despite this being the trivially gapped paramagnetic phase, the gap remains 
anomalously small.}
\label{fig:Bxym2z}    
\end{figure}

\noindent
{\bf Field-induced chiral spin liquid.}  If the field components satisfy the 
conditions $B_x \neq 0$, $B_y \neq 0$, $B_z \neq 0$, and $B_c \neq 0$ in such 
a way that $\mathcal C \neq 0$, then a chiral spin liquid can be induced. 
In Fig.~\ref{fig:Bxyz} we show the spinon dispersion for the case $\pmb B 
\parallel \hat c$, which ensures that $\mathcal C = 2$. When the field 
is sufficiently strong that magnetic order has been suppressed 
[Fig.~\ref{fig:Bxyz}(a)], which is a first-order transition (Fig.~4(b) of 
the main text), the resulting gapped phase is a chiral spin liquid. We draw 
attention to the fact that the gap in question is between the second and 
third bands (half-filling) in all panels of Fig.~\ref{fig:Bxyz}, whereas 
the large gap visible between the first and second bands in this case is not 
relevant. As the field is increased to a critical value, the spinon band gap 
closes at two symmetrical points in the Brilloun zone [Fig.~\ref{fig:Bxyz}(b)]. 
A further increase in field causes the gap to reopen [Fig.~\ref{fig:Bxyz}(c)] 
in the topologically trivial gapped phase.
 
\begin{figure}[b]
\includegraphics[width=8.5cm]{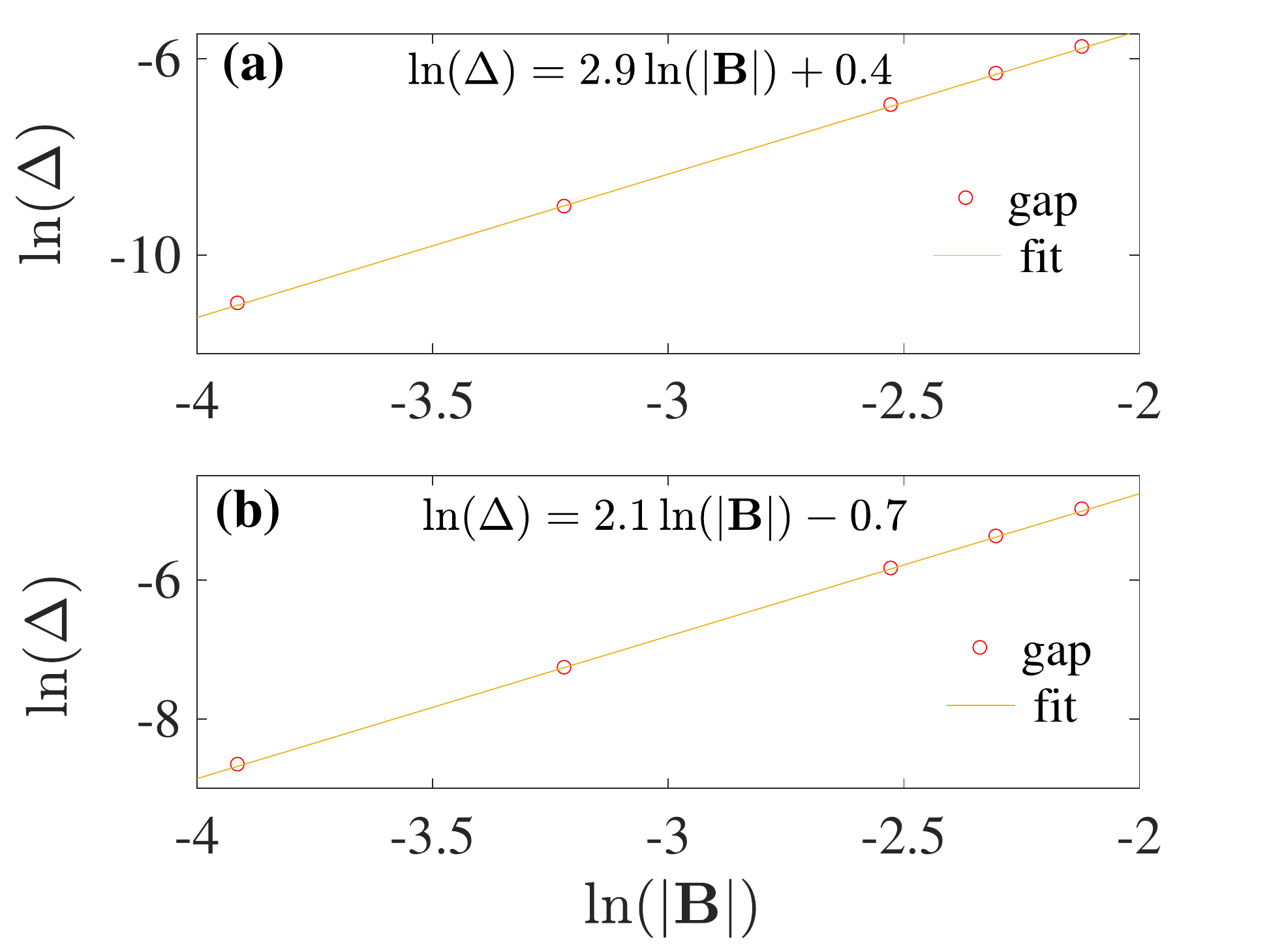}
\caption{Scaling of the energy gap, $\Delta$, in the spinon dispersion, 
shown as a function of the field strength, $|\pmb B|$, on logarithmic axes 
for two field orientations with $B_c = 0$. (a) $\pmb B \parallel (\hat x + 
\hat y - 2 \hat z)$. (b) $\pmb B \parallel (\hat x + 2\hat y - 3\hat z)$.}
\label{fig:powgap}    
\end{figure}

\noindent
{\bf Trivial paramagnetic phase.} If field direction is such that the system 
obeys none of the special symmetries classified in the main text, i.e.~$B_x 
\neq 0$, $B_y \neq 0$, $B_z \neq 0$, and $B_c \neq 0$ with $\mathcal C = 0$, 
then a gapped paramagnetic phase is induced when the field is strong enough 
to suppress the magnetic order. In this case, there is only one phase 
transition (Fig.~4(c) of the main text) and the gapped state is connected 
directly to the fully polarized phase, meaning that it is topologically 
trivial. In this phase the spinons are confined, making the excitations 
of the system bosonic. 

Also in this phase are the special cases when one of the components 
$B_x$, $B_y$, $B_z$, or $B_c$ is zero, i.e.~for field directions on the 
circles shown in Fig.~1(a) of the main text. This situation also has only 
one phase transition, but, as stated in the main text, the gap of the 
field-induced paramagnetic phase does not open linearly, following instead 
an algebraic form. Concerning the size of this gap, in Fig.~\ref{fig:Bxym2z} 
we show the example $\pmb B \parallel (\hat x + \hat y - 2 \hat z)$, where 
the gap scales with the field according to $\Delta \propto |B|^{2.9}$, making 
the gap for $g \mu_B |\pmb B|/|K| = 0.24$ only $\Delta \simeq 0.07|K|$.
Concerning the algebraic functional form of this gap, we have conducted 
careful numerical studies to investigate its scaling with the intensity of 
the applied field (in this analysis we set $M = 0$ and consider very low 
fields). Figure \ref{fig:powgap} illustrates two different cases: if $\pmb B 
\parallel (\hat x + \hat y - 2 \hat z)$, then the gap scales with the field 
according to $\Delta \propto |\pmb B|^{2.9}$; if $\pmb B \parallel (\hat x
 + 2\hat y - 3\hat z)$, then $\Delta \propto |\pmb B|^{2.1}$. By contrast, 
when all of the masses ($B_x,B_y,B_z,B_c$) are nonzero, regardless of whether 
or not the total Chern number is zero, then the gap scales linearly with 
$|\pmb B|$. 

\section{Field orientation and quantum phase transitions} 

It is illustrative to consider the stability of the different spin-liquid 
states as a function not only of the field strength but also of the angle at 
which the field is applied. For the special points at which the system has a 
gapless, four-cone dispersion, this Dirac QSL is not stable: for appropriate 
field strengths, any change of field angle will cause the system to open a gap. 
In this sense the U(1) Dirac QSL is different from the Z$_2$ Kitaev QSL, which 
is protected against such small angle changes by the finite vison gap. Along 
the lines where a change of field angle causes the Chern number to change from 
$\mathcal C = \pm 2$ to 0, the spinon gap closes and the dispersion has two 
Dirac cones; as discussed in the main text, these are quantum phase transitions 
between states of confined and deconfined spinons. Because of the finite field 
strength, these transition lines do not lie strictly on the large circles of 
Fig.~1 of the main text, but are instead deformed slightly towards the chiral 
QSL phase, whose regime of stability is therefore a little smaller than Fig.~1 
would indicate. The Dirac QSL (four-cone dispersion) points actually mark the 
meeting of four phases, two with $\mathcal C = 0$ and one each with $\mathcal 
C = \pm 2$, and their positions as transition points remain unchanged for any 
field strength. 

\end{document}